\documentclass[pra,superscriptaddress,aps,10pt,reprint,nofootinbib]{revtex4-1}

\usepackage[utf8]{inputenc}
\usepackage[T1]{fontenc}
\usepackage[english]{babel}

\usepackage{graphicx}

\usepackage{amssymb,amstext}
\usepackage{amsthm}
\usepackage{amsmath}
\usepackage{amsfonts}
\usepackage{mathtools}

\usepackage[colorlinks=true, allcolors=blue]{hyperref}
\usepackage[all]{hypcap}  % for images: anchor the image, not the caption

\usepackage{braket}
\usepackage{xcolor}
\usepackage[normalem]{ulem}

\renewcommand{\v}[1]{\vec{\boldsymbol{#1}}}
\newcommand{\vu}[1]{\hat{\boldsymbol{#1}}}
\newcommand{\dyad}[1]{|#1\rangle\langle#1|}

\begin{document}
\title{Alignment of $s$-state Rydberg molecules in magnetic fields}
\date{\today}
\author{Frederic Hummel}
\email{frederic.hummel@physnet.uni-hamburg.de}
\affiliation{Zentrum für Optische Quantentechnologien, Universität Hamburg, Luruper Chaussee 149, 22761 Hamburg, Germany}
\author{Christian Fey}
\affiliation{Zentrum für Optische Quantentechnologien, Universität Hamburg, Luruper Chaussee 149, 22761 Hamburg, Germany}
\author{Peter Schmelcher}
\affiliation{Zentrum für Optische Quantentechnologien, Universität Hamburg, Luruper Chaussee 149, 22761 Hamburg, Germany}
\affiliation{Hamburg Centre for Ultrafast Imaging, Universität Hamburg, Luruper Chaussee 149, 22761 Hamburg, Germany}

\begin{abstract}
We unravel some peculiar properties of ultralong-range Rydberg molecules formed by an $s$-state $^{87}$Rb Rydberg atom and a corresponding ground-state atom whose
electronic orbitals are spherically symmetric and therefore should not be influenced by the presence of weak magnetic fields. 
However, the electron-atom interaction, which establishes the molecular bond, is under certain conditions subject to a sizable spin-orbit coupling and, hence, sensitive to the magnetic field.
This mechanism can be harnessed to counterintuitively align the $s$-state molecules with respect to the field axis.
We demonstrate this by analyzing the angular-dependent Born-Oppenheimer potential energy surfaces and the supported vibrational molecular states.
Our predictions open interesting possibilities for accessing the physics of relativistic electron-atom scattering experimentally.
\end{abstract}

\maketitle

\section{Introduction}
The extraordinary properties of ultra-long-range Rydberg molecules (ULRM) such as their enormous size and large permanent electric dipole moments (PEDM) have stirred increasing interest in their study. 
Two main molecular species can be distinguished: nonpolar molecules with low electronic angular momentum, e.g., $s$-state or $d$-state ULRM, and polar ones with high angular momentum. 
The latter are called trilobite and butterfly molecules after their electronic wave function shaped by $s$- and $p$-wave interactions with the ground state atom \cite{greene_creation_2000, hamilton_shape-resonance-induced_2002, Chibisov2002}. \\
In recent years, a multitude of spectroscopic experiments have probed ULRM from their first detection in rubidium \cite{Bendkowsky2009}, cesium \cite{Tallant2012}, and strontium \cite{DeSalvo2015}, their coherent excitation \cite{Butscher2010}, lifetime measurements \cite{Butscher2011, Camargo2016, Whalen2017}, the observation of PEDM \cite{li_homonuclear_2011, Booth2015} and angular momentum couplings \cite{Anderson2014a}, to proving the existence of polymers \cite{Bendkowsky2010, gaj_molecular_2014} and controlling ULRM in external fields \cite{krupp_alignment_2014, Gaj2015, niederprum_observation_2016}. 
Additionally, several applications have been proposed and implemented such as probing lattice gases via ULRM \cite{Manthey2015} and performing remote spin flips \cite{niederprum_rydberg_2016}, as well as the study of ultracold chemical reactions \cite{Schlagmuller2016x} and electromagnetically induced transparency \cite{Mirgorodskiy2017}. 
Moreover, they provide the possibility to realize Rydberg and ionic impurities in Bose-Einstein condensates \cite{balewski_coupling_2013, Wang2015, Camargo2017, Eiles2016, Schmidt2016, Schmidt2018, Kleinbach2018}, and to study ion-atom interaction \cite{Schmid2018, Engel2018} as well as to employ optical Feshbach resonances in order to tune atom-atom interaction \cite{Sandor2017,Thomas2017}. \\
The molecular binding originates from $s$- and $p$-wave scattering of the Rydberg electron off a neutral ground-state atom \cite{Fermi1934, omont_theory_1977}. 
Therefore, ULRM serve as a unique tool to characterize electronic scattering channels \cite{Sassmannshausen2015, Boettcher2016} and resonances at very low electronic energies \cite{Schlagmuller2016, MacLennan2018}.
Specifically, they provide an excellent platform for experimentally testing \textit{ab initio} calculations of $s$- and $p$-wave scattering phase shifts \cite{Fabrikant1986, Bahrim2000, Bahrim2001, Eiles2018}.
In particular, in rubidium and cesium the electron-neutral scattering possesses a $p$-wave shape resonance, which can be understood as a short-lived, metastable, anionic state of the electron-neutral system \cite{Johnston1982, Johnston1995, Scheer1998}. 
A higher level theory that takes relativistic spin-orbit coupling of the two valence electron's spin with the their angular momentum relative to the ground-state atom's core into account has been provided \cite{Khuskivadze2002, Markson2016, Eiles2017}. \\
In this paper, we demonstrate that the relativistic spin-orbit interaction paired with a weak magnetic field gives rise to the surprising feature of alignment of $s$-state ULRM.
For $d$-state ULRM, alignment in magnetic fields has been demonstrated and stems from the nonzero orbital angular momentum of the electronic state \cite{krupp_alignment_2014, Hummel2017}.
However, for $s$-state, i.e., zero angular momentum, ULRM such an alignment process is unexpected, due to the isotropy of the electronic state, since $l$ mixing is strongly suppressed for nonpolar ULRM (see the appendix).
Indeed, for $s$-state ULRM, the alignment originates from the interaction of the electronic spin degrees of freedom with the spatial degrees of freedom within the $p$-wave scattering.
A pictorial illustration of the mechanism is presented in Fig.\@ \ref{Mechanism}:
\begin{figure}
\centering
\includegraphics[width=1\columnwidth]{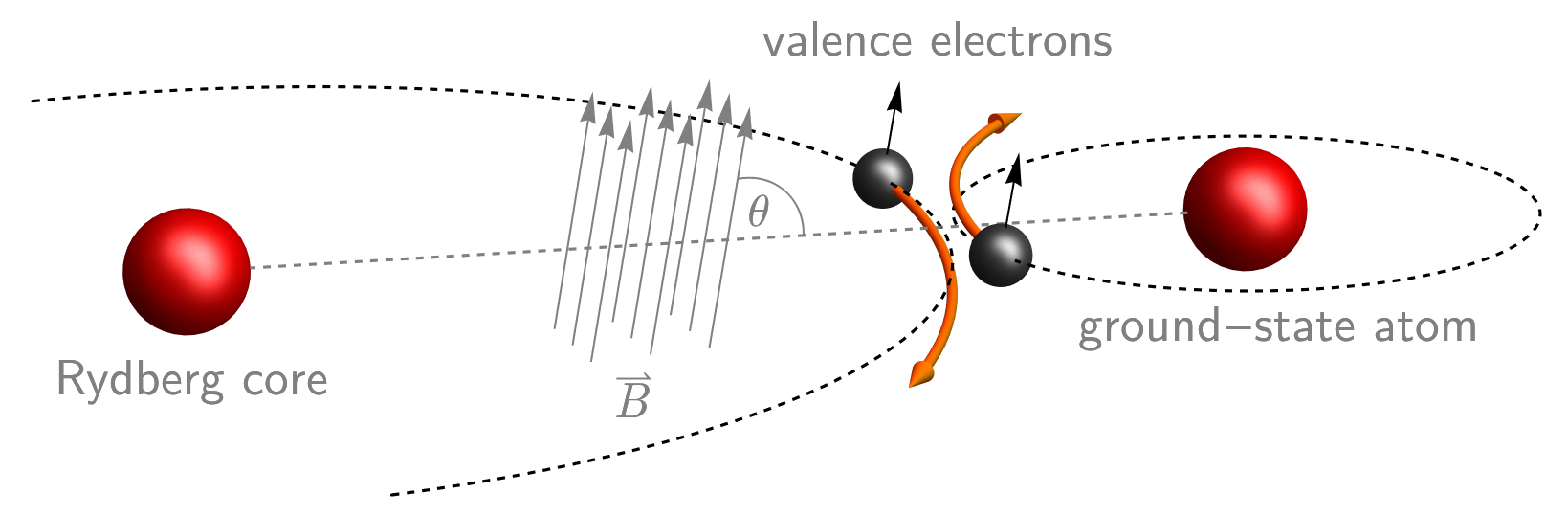}
	\caption{Pictorial representation of the alignment mechanism of the Rydberg electron and the ground-state atom. The magnetic field (gray arrows) tends to orient the electronic spins (black arrows). Because of spin-orbit interaction, the relative angle with regard to the internuclear axis $\theta$ influences the scattering process (orange arrows) and leads, consequently, to angular-dependent potential energy surfaces. \label{Mechanism}}
\end{figure}
In a first step, the magnetic field couples to the electronic spins and tends to align them.
In a second step, the spin-orbit coupling modifies the electron-atom interaction in an angular-dependent fashion. \\
We analyze this alignment by determining the angular-dependent Born-Oppenheimer potential energy surfaces (PES) as well as energies and wave functions of bound vibrational states.
To provide a more intuitive physical picture of the underlying mechanism, we present an approach that considers the presence of $p$-wave shape resonances and the magnetic field perturbatively and reproduces the polar potential energy curves of the underlying \textit{ab initio} calculations. 
We focus here on rubidium molecules.
However, our approach applies also to ULRM of other atomic species that possess a $p$-wave shape resonance such as cesium. \\

\section{Setup and Interactions}
\begin{figure}
\centering
\includegraphics[width=0.65\columnwidth, angle=270]{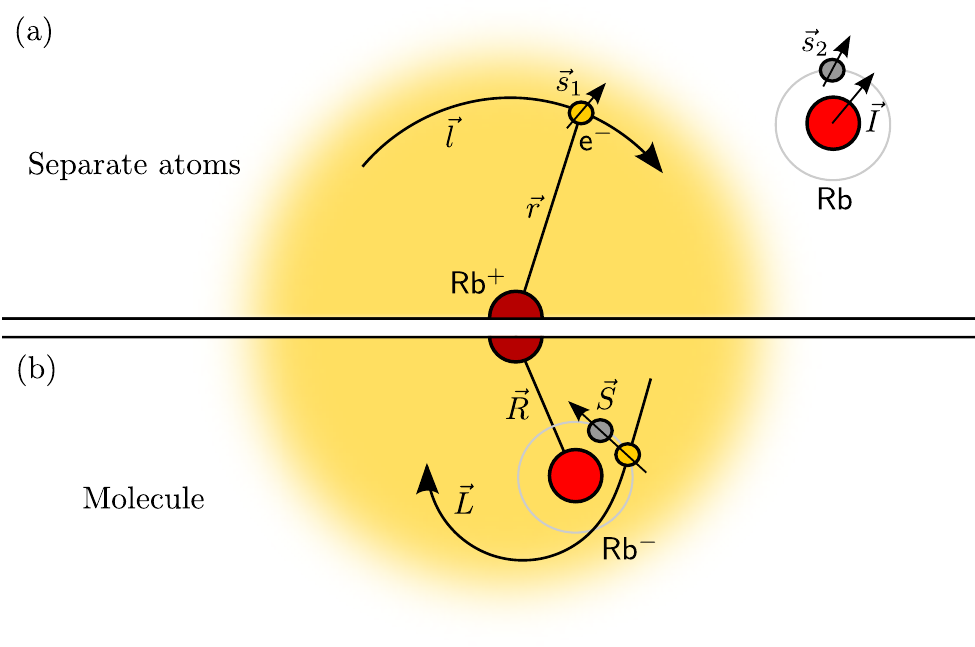}
	\caption{Sketch of the molecule and its spin degrees of freedom in two different representations. 
	In panel (a), the ground-state atom with nuclear spin $\v{I}$ and electronic spin $\v{s}_2$ is located outside the electronic cloud of the Rydberg atom (yellow shading). The Rydberg atom's core is located in the center and the Rydberg electron at position $\v{r}$ carries orbital angular momentum $\v{l}$ and spin $\v{s}_1$. 
	In panel (b), the ground-state atom is located inside the cloud at position $\v{R}$. Here, the electronic spins are coupled to the total electronic spin $\v{S}$ which again couples to the electron pair's angular momentum relative to the ground-state atom's core $\v{L}$. \label{Sketch}}
\end{figure}

Our diatomic ULRM consists of a Rydberg and a ground-state rubidium atom.
The closed shell core electrons of the Rydberg atom are taken into account by quantum defects of the Rydberg energy levels obtained by experimental observation \cite{Li2003, Han2006}. 
The polarizable ground-state atom with hyperfine structure \cite{Arimondo1977} at relative position $\v{R}$ to the Rydberg core acts as a perturber to the Rydberg electron's wave function (see Fig.\@ \ref{Sketch} for a sketch including all relevant spin degrees of freedom). 
The interaction of the ground-state atom with the Rydberg electron can be described via a generalized Fermi pseudopotential (we use atomic units unless stated otherwise),
\begin{equation}
V(\v{R},\v{r}) = \sum_{\beta} \frac{(2L+1)}{2} a({}^{2S+1}L_J,k) \frac{\delta(x)}{x^{2(L+1)}} \dyad{\beta},
\label{eqn:V_pseudo}
\end{equation}
which takes into account partial waves up to $p$-wave interactions that are proportional to the respective spin-dependent scattering lengths \cite{Khuskivadze2002, Eiles2017}.
This includes six scattering channels labeled with angular momentum quantum numbers in the frame of the ground-state atom, ${}^{2S+1}L_J = \{ {}^1S_0, {}^3S_1, {}^1P_1, {}^3P_0, {}^3P_1, {}^3P_2 \}$, where $S$ is the total spin of the valence electron pair, $L$ is the electron pair's orbital angular momentum relative to the ground-state atom, $J$ is the total electronic angular momentum arising from $\v{L}\cdot\v{S}$ coupling, and $M_J$ is the total electronic angular momentum projection onto the internuclear axis. 
The scattering states are labeled as $\ket{\beta}=\ket{(L,S)J,M_J}$, $x=|\v{r}-\v{R}|$ is the distance between the Rydberg electron and the neutral ground-state atom, and $a({}^{2S+1}L_J,k)$ is the energy-dependent scattering length associated to the respective channel. 
The wave number $k$ is approximated via the semi-classical electron energy $\tfrac{1}{2}k^2(R)=\tfrac{1}{R}-\tfrac{1}{2n_h^2}$, where $R$ is the distance between the two atoms and $n_h$ is the principal quantum number of the closest hydrogenic manifold. 
Although the contact potential in (\ref{eqn:V_pseudo}) leads to formally ill-behaved solutions, it approximates the electron-atom interaction reasonably well, when employed in a restricted basis set approach \cite{Du1987, Fey2015, Eiles2016}, in very good agreement with experiments \cite{Bendkowsky2009}. 
Alternative approaches are based on finite-range pseudopotentials, Green's function methods, or $R$-matrix techniques \cite{Khuskivadze2002,hamilton_shape-resonance-induced_2002,Tarana2016}.
The ground-state atom's hyperfine structure is considered via $H_\text{HF} = A \v{s}_2 \cdot \v{I}$, where $\v{s}_2$ is the ground-state atom's electronic spin, $\v{I}$ is its nuclear spin, with $I=\tfrac{3}{2}$ for rubidium, and $A=3.417\ \text{GHz}$ is rubidium's hyperfine constant.
The effect of a magnetic field is included via the corresponding Zeeman term and couples linearly to the electronic angular momenta
\begin{equation}
H_{\text{B}} = \v{B} \cdot \left( \v{S} + \v{l}/2 \right),
\label{eqn:HB}
\end{equation}
where $\v{B}$ is given in units of $2.35\times10^9$ Gauss and $l$ is the Rydberg electrons's orbital angular momentum relative to the Rydberg core, which is nonzero for states other than the $s$ state. \\
The combined electronic Hamiltonian $H=H_\text{Ryd}+ V+ H_\text{HF}+ H_\text{B}$ \cite{Eiles2017, Hummel2017} is diagonalized within the Born-Oppenheimer approximation, yielding the adiabatic PES, which depend parametrically on the distance between the Rydberg core and the ground-state atom. 
Indeed, due to the azimuthal molecular symmetry corresponding to rotations around the magnetic field axis, the PES depend only on the internuclear distance $R$ and the polar angle $\theta$ between the magnetic field axis and the internuclear axis. 
Our basis includes the closest hydrogenic manifold, which lies energetically above the considered $s$ state (Fig.\@ \ref{Radial}), with all Rydberg electronic angular momentum states that have a maximal projection onto the internuclear axis of $m_{\text{max}}=3/2$, and all ground-state atomic hyperfine states. 
Further increase of $m_{\text{max}}$ has only minor influence on our results (see the appendix, cf.\@ \cite{Kleinbach2017}). \\

\section{Results}
\begin{figure}
	\centering
	\includegraphics[width=1.0\columnwidth]{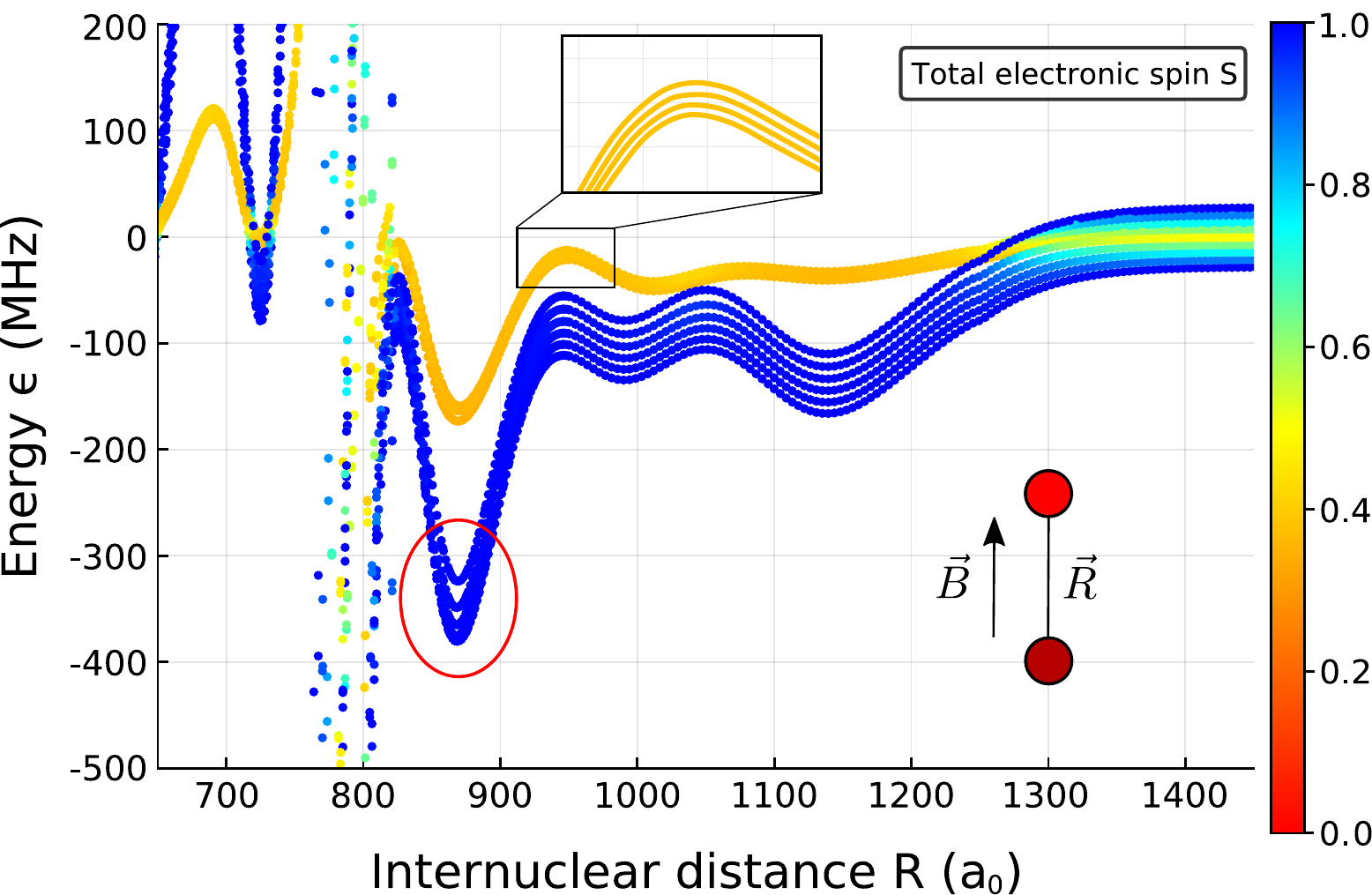}
	\caption{Radial cut of the PES for a rubidium $28s$ state in a magnetic field of $B=10$ G parallel to the internuclear axis ($\theta=0$) for the ground-state atom in a hyperfine $F=2$ state. The color code indicates the expected total electronic spin. A pure triplet curve with sixfold multiplicity is visible as well as a mixed singlet-triplet curve with fourfold multiplicity (inset). In the deepest potential well around $R=870 \ \text{a}_0$ (see encircled region), which lies close to the $p$-wave shape resonance, the Zeeman splitting is not equidistant. \label{Radial}}
\end{figure}

\begin{figure}
	\centering
	\includegraphics[width=1\columnwidth]{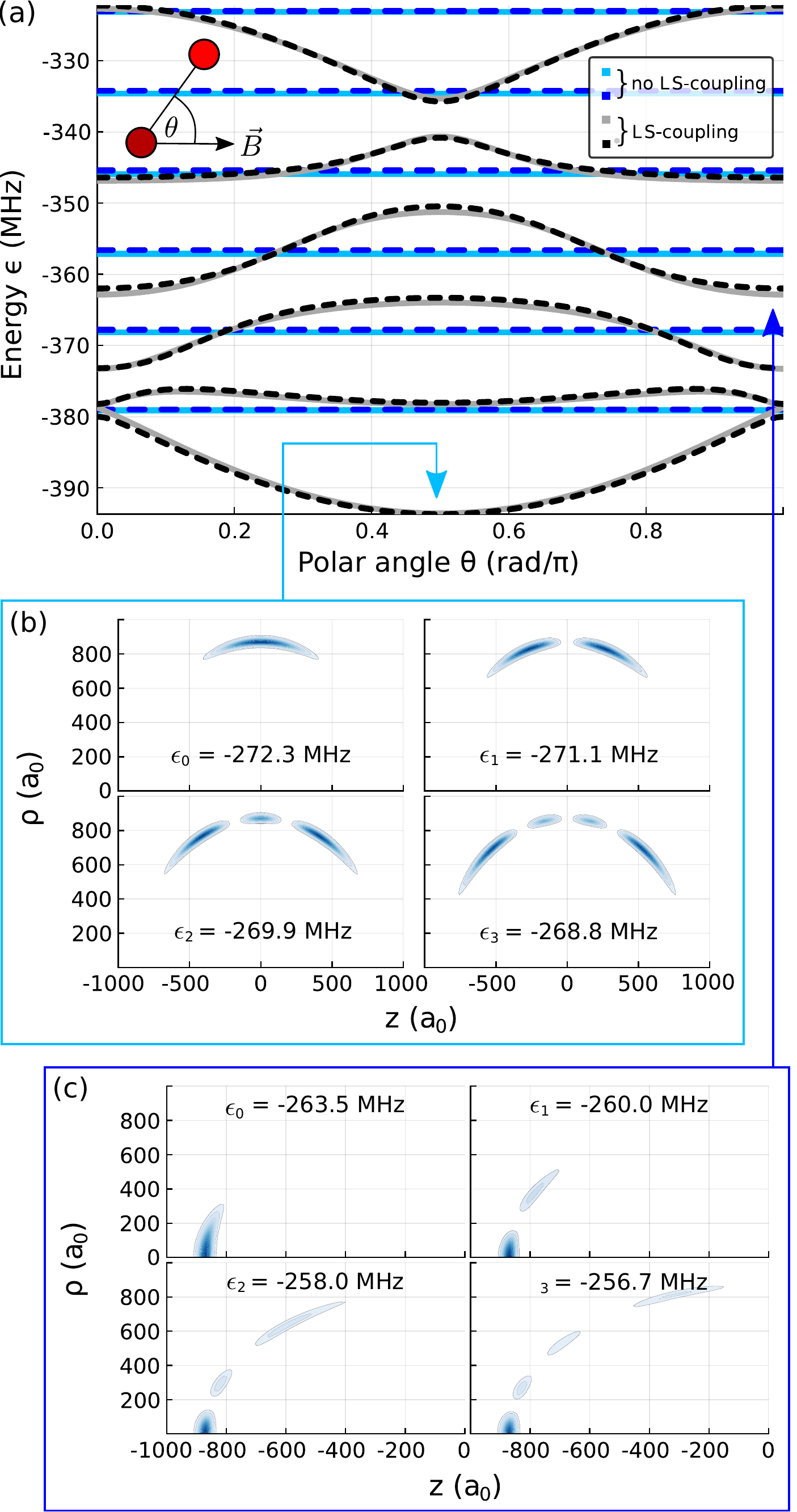}
	\caption{(a) Angular cut of the PES for a rubidium $28s$ state in a magnetic field of $B=10$ G at the radial position $R=870 \ \text{a}_0$ matching the position of the potential well close to the $p$-wave shape resonance (see Fig.\@ \ref{Radial}). The numerical result within exact diagonalization of the electronic Hamiltonian (light blue and gray lines) is compared to a reduced model only taking the six relevant states of the $K=5/2$ multiplet into account (blue and black dashed lines). Both approaches are considered with equal (light blue, blue) and distinguishable (gray, black) triplet $p$-wave scattering lengths. 
	[(b), (c)] Probability densities in cylindrical coordinates of vibrational states $u(\rho,z)$ supported by the $m_F=-2$ and $m_F=0$ dominated states, respectively. The ground-state atom localizes in the perpendicular (parallel) configuration relative to the magnetic field axis. The eigenstate energy $\epsilon_\nu$ is provided relative to the unperturbed $28s$, $F=2$ atomic energy. The vibrational ground state features an alignment of $\langle\cos^2\theta\rangle = 0.04$ (b) and 0.98 (c). \label{Model}}
\end{figure}

Figure \ref{Radial} depicts a radial cut of the PES for rubidium $28s$ (sPES) and $F=2$ in a magnetic field of $B=10$ G that is aligned parallel to the internuclear axis. 
The presence of the neutral ground-state atom in the Rydberg electronic cloud leads to an oscillatory potential curve typical for ULRM. 
Illustratively speaking, the emerging potential wells act as traps for the neutral ground-state atom and lead to radially localized bound states between the Rydberg atom and the ground-state atom, thus forming a molecule. 
At distances $R\approx800 \, \text{a}_0$, where $\text{a}_0$ is the Bohr radius, the kinetic energy of the Rydberg electron matches the energy of the $p$-wave shape resonances, which leads to a steep crossing of the butterfly PES.
Sufficiently far from the resonance, the PES can be characterized by their total electronic spin character, which is shown in Fig.\@ \ref{Radial} as a color code.
Deep potential wells occur when the electronic spins are aligned to a triplet state ($S=1$), and shallow wells when the spins are antialigned in a mixed singlet-triplet state. 
The presence of the magnetic field splits the otherwise degenerate curves and reveals their multiplicity, which in the case of the $F=2$ hyperfine state is six for the triplet curve and four for the mixed curve. \\
Typically within a region of $100 \, \text{a}_0$ around the crossing with the butterfly curve (at $R\approx800 \, \text{a}_0$), we find a significant dependence of the sPES on the polar angle $\theta$. 
An angular cut of the sPES at the radial position of the deepest potential well at $R=870 \ \text{a}_0$ is shown in gray in Fig.\@ \ref{Model}(a). 
The emerging angular potential wells are deep enough to support aligned molecular states for which three equilibrium angles are possible.
Both the energetically lowest and highest visible sPES allow for molecules oriented perpendicular to the magnetic field axis ($\theta=\tfrac\pi2$), whereas all other visible sPES allow for parallel and antiparallel alignment ($\theta=\{0,\pi\}$).
The second lowest visible sPES shown in gray in Fig.\@ \ref{Model}(a) constitutes an exception to this and is mostly flat not allowing any alignment.
The angular dependence of the sPES is only present if the spin-orbit interaction is included in the description and the $p$-wave interaction is sufficiently strong.
When choosing equal scattering lengths for all three triplet $p$-wave channels and therefore effectively eliminating the relativistic spin-orbit coupling in the scattering event, the polar-angle dependence vanishes, which is shown in light blue. \\
To describe vibrational states belonging to the PES, it is beneficial to introduce cylindrical coordinates, i.e., $\v{R}= \rho (\cos\phi \, \vu{e}_x + \sin\phi \, \vu{e}_y)+ z\, \vu{e}_z$, with the $z$ axis pointing along the direction of the $\v{B}$ field. 
In this case, stationary states can be expressed as $\chi(\v{R})=u(\rho,z)  \text{exp}(-i \mu \phi)/\sqrt{\rho}$ with a good azimuthal angular momentum quantum number $\mu$. 
The states $u(\rho,z)$ are governed by the Hamiltonian \cite{krupp_alignment_2014, Hummel2017}
\begin{equation}
H_{\text{vib}} = -\frac{1}{M} (\partial_\rho^2 + \partial_z^2) + \frac{\mu^2 - 1/4}{M\rho^2} + \epsilon_i(\rho,z),
\label{eqn:Hvib}
\end{equation}
where $M$ is the atomic mass of $^{87}$Rb and $\epsilon_i(\rho,z)$ are the PES.
We focus exclusively on $\mu=0$ and obtain the eigenstates of (\ref{eqn:Hvib}) via a two-dimensional finite difference method.
For the energetically lowest curve in Fig.\@ \ref{Model}(a), which represents the spin-polarized $F=2$, $m_F=-2$ state, the probability densities of the vibrational ground state and the first three excited states are provided in Fig.\@ \ref{Model}(b).
They exhibit an energy spacing of $1$ MHz and the vibrational ground state features an alignment of $\langle\cos^2\theta\rangle = 0.04$.
The third gray curve from the bottom in Fig.\@ \ref{Model}(a) corresponds to an $s$ state dominated by  $F=2$, $m_F=0$ contributions and possesses potential wells around $\theta=0$ and $\theta=\pi$.
The probability densities of the corresponding vibrational states can be seen in Fig.\@ \ref{Model}(c) and exhibit an energy spacing of $2$ MHz, while the vibrational ground state features an alignment of $\langle\cos^2\theta\rangle = 0.98$. \\

\section{Discussion}
In order to develop a more intuitive picture of our alignment mechanism, we reduce the complexity of the electronic Hamiltonian by means of a perturbative approach.
To reproduce the angular cuts of the PES, it is sufficient to restrict the Hilbert space to a subspace with $l=0$ and a constant quantum number $K=|\v{S}+\v{I}|=|\v{s}_1+\v{F}|$.
For $s$-state ULRM, $K$ 
represents the molecular system's total angular momentum and is approximately conserved (up to admixture of $l>0$ states).
This is reflected by the multiplicity of the PES in Fig.\@ \ref{Radial}.
The extreme values for $K=\{\tfrac12, \tfrac52\}$ correspond to pure triplet ($S=1$) states, whereas $K=\tfrac32$ states have mixed singlet and triplet character. 
In this subspace, the Hamiltonian reads
\begin{equation}
H(\theta) = c_\text{s} + c_\text{p} \sum_{J=0}^{2} V_\text{p}(J) + H_B,
\label{eqn:H_model}
\end{equation}
where $c_\text{s}$ and $c_\text{p}$ are parameters which are adjusted to reproduce the results of the field-free exact diagonalization. 
$c_\text{s}$ corresponds to the overall offset due to the dominant $s$-wave interaction and $c_\text{p}$ controls the admixture of the $p$-wave interaction.
The $p$-wave interactions $V_\text{p}(J)$ depend on the total electronic angular momentum $J$ relative to the ground-state atom at $\v{R}$, while $H_B$ couples the electronic spins to the $\v{B}$-field axis. 
The resulting PES will, however, only depend on the relative angle $\theta$ between $\v{R}$ and $\v{B}$. 
Without loss of generality we can therefore specify  $\v{R}= R \vu{e}_z$ and $\v{B}= B (\cos \theta \, \vu{e}_z + \sin \theta \, \vu{e}_x$).
In the $l=0$ subspace, the $p$-wave interaction terms in  (\ref{eqn:V_pseudo}) simplify then to
\begin{equation}
V_\text{p}(J) = a({}^3P_J, k) \sum_{M_S}|C_{10, 1M_S}^{JM_S}|^2 \dyad{M_S},
\label{eqV}
\end{equation}
where $M_S$ are the three possible projections of $S=1$ onto the internuclear $z$ axis and $C_{LM_L, SM_S}^{JM_J}$ is a Clebsch-Gordan coefficient describing the coupling of $\v{S}$ and $\v{L}$ to $\v{J}$.
The impact of the magnetic field on $s$ states reduces from (\ref{eqn:HB}) to
\begin{equation} 
H_B = B \left( S_z \cos \theta + S_x \sin \theta \right),
\label{eqn:HB_for_l0}
\end{equation}
with the spin matrices for $z$ and $x$ direction $S_{z,x}$.
The simple structure of (\ref{eqV}) and (\ref{eqn:HB_for_l0}) reveals the nature of the alignment mechanism: 
If the scattering lengths for the $p$-wave channel $a({}^3P_J, k)$ are equal, the generator of rotations $\theta$ in (\ref{eqn:HB_for_l0}), here $S_y$, commutes with $V_\text{p}$, and the eigenvalues of $H(\theta)$ are independent of $\theta$. 
However, in the case of different $p$-wave triplet scattering lengths an angular dependence emerges.
To compare the model (\ref{eqn:H_model}) to the full Hamiltonian, we diagonalize the matrix
$\bra{KM_K} H(\theta) \ket{KM'_K}$, 
where $M_K$ is the projection of $K$ onto the internuclear axis. \\
Figure \ref{Model}(a) shows the sPES of the model in the case of $K=5/2$ for a magnetic field of $B=10$ G, which corresponds the curves circled in red in Fig.\@ \ref{Radial}.
For distinguishable (equal) scattering lengths, shown in black (blue) the model recovers the results of the exact diagonalization, which are shown in gray (light blue). 
For larger magnetic fields, when $H(\theta)$ is dominated by $H_B$, the different PES separate further while the depth of the potential wells does not increase. 
For smaller magnetic fields, when $H(\theta)$ is dominated by $V_\text{p}(J)$, the PES are structured in pairs of equal absolute total angular momentum projection $|M_K|$. 
The $\theta$-dependent influence of the interaction matrices $V_\text{p}(J)$, which represent the three different triplet $p$-wave scattering channels, on the eigenvalues of $H(\theta)$ corresponds to the physical picture that the spin-orbit interaction introduces a spatial degree of freedom and consequently angular anisotropy to the otherwise isotropic molecular system. \\

\section{Conclusion}
We predict the possibility of molecular alignment of $s$-state ultra-long-range Rydberg molecules in a homogeneous magnetic field.
In contrast to $d$-state ULRM that can be aligned in magnetic fields due to the nonzero orbital angular momentum of the Rydberg state, the alignment of $s$-state ULRM has a completely different origin.
We attribute the effect to the interplay of a magnetic field and the spin-orbit coupling of the electron-atom interaction.
For the alignment to occur in $s$ states, it needs both the $\v{L}\cdot\v{S}$- and the strong $p$-wave interaction.
The alignment mechanism can be interpreted by means of a reduced interaction model in a comparably low-dimensional Hilbert space. 
This novel degree of control of the otherwise isotropic $s$ states of ULRM is clearly within reach of current experimental efforts.
Although signatures in measured electric dipole moments of butterfly ULRM indicate the presence of $\v{L}\cdot\v{S}$ couplings in ULRM \cite{Eiles2016}, a clear 
experimental confirmation is still missing. 
An experimental observation of the here proposed alignment of $s$-state ULRM would be perfectly suited for such a purpose. \\
Beyond this work, it would be beneficial to study the influence of higher angular momentum states on the alignment, for instance in trilobites or $p$-state molecules, with the butterfly molecules being a prominent candidate to show effects of spin-orbit interaction. 
Furthermore, similar effects should occur in polyatomic ULRM and are expected to lead to novel forms of angular dependent three-body interactions.
ULRM are an exceptionally suitable environment to experimentally study low-energy electronic scattering. 
In particular, the study of spin-orbit interaction effects provides unique possibilities to experimentally characterize relativistic electron-atom $p$-wave scattering and its underlying resonances. 
\begin{acknowledgements}
F.H.\@ and P.S.\@ acknowledge support from the German Research Foundation (DFG) within the priority program "Giant Interactions in Rydberg Systems" (SPP 1929 GiRyd). C.F.\@ gratefully acknowledges a scholarship from the Studienstiftung des deutschen Volkes.
\end{acknowledgements}
\appendix*
\section{$l$ MIXING OF MOLECULAR STATES} \label{lmix}
\begin{figure}
\centering
\includegraphics[width=1\columnwidth]{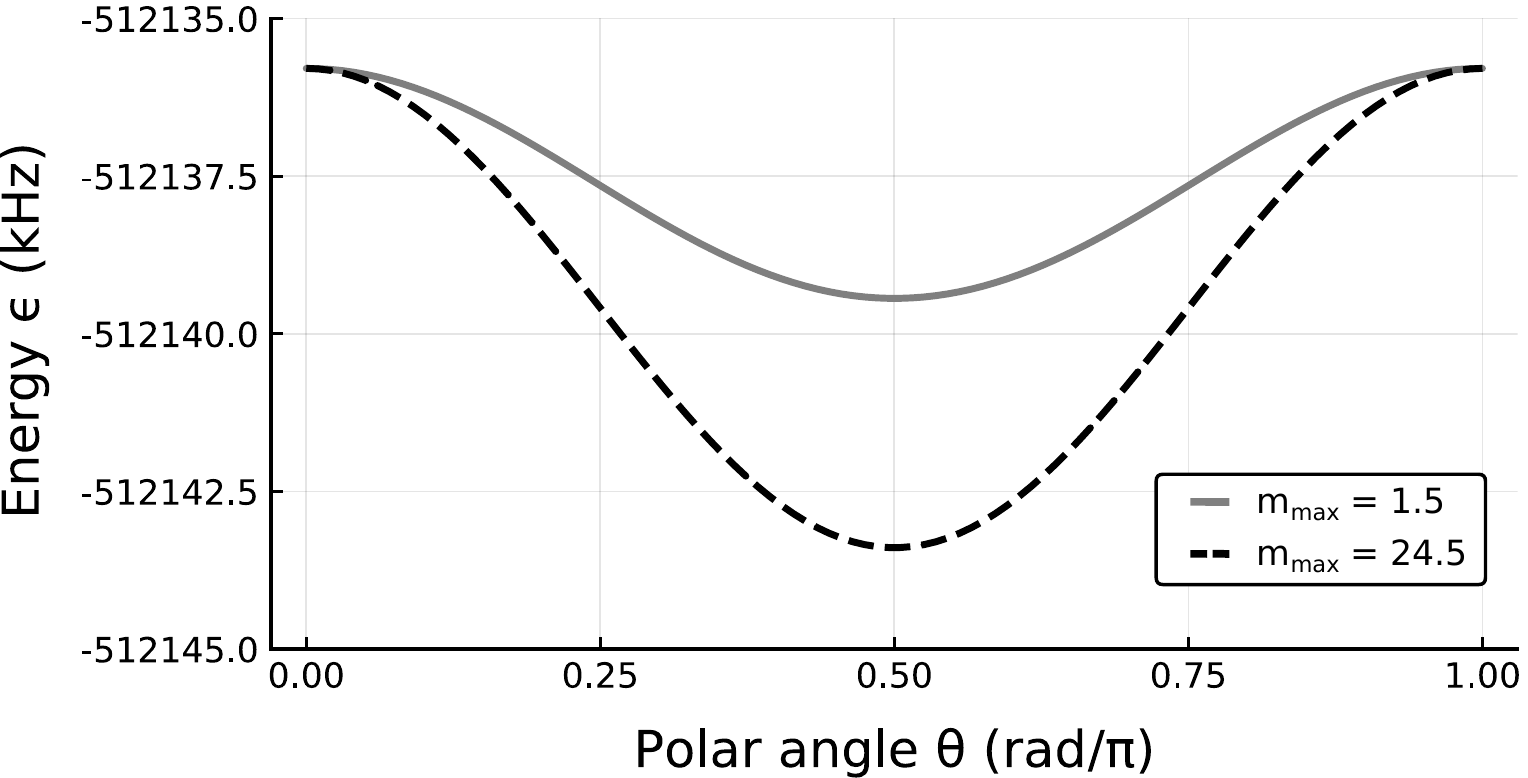}
	\caption{The same setup as in Fig.\@ \ref{Model}(a) neglecting hyperfine structure and $\v{L}\cdot\v{S}$ coupling.
	The results for different maximal total angular momentum projections of the Rydberg electron $m_{\text{max}}$ are shown.
	The inclusion of more projection states makes the an angular dependency more pronounced. The maximal energetic difference of different polar angles between the internuclear axis and the magnetic field is on the order of a few kHz. \label{Angular}}
\end{figure}
In general, the interaction of the Rydberg electron with the ground-state atom leads to $l$ mixing. 
However, due to the noninteger quantum defect of atomic states with low orbital angular momentum $l<3$ and resulting energetical detuning off the degenerate hydrogenic manifold, this process is strongly suppressed for Rydberg $s$ states discussed here.
To quantitatively examine the impact of $l$-mixing on the alignment process, we neglect the hyperfine structure of the ground-state atom in the absence of $\v{L}\cdot\v{S}$ coupling. 
This approach makes it numerically feasible to consider all total angular momentum projection quantum numbers of the Rydberg electron $m_j$ up to $m_{\text{max}}=n-1+\tfrac{1}{2}=24.5$. 
Figure \ref{Angular} shows an angular cut of the derived PES for the $28s+5s$ molecule for such a calculation (black, dashed line) at the radial position of the inner potential well shown in Fig.\@ \ref{Radial} at $R=870 \ \text{a}_0$ and compares it to a calculation where $m_{\text{max}}=1.5$ (gray, solid line). 
While the inclusion of all projection states makes the effect more pronounced, an angular dependence is visible on the scale of a few kHz, which is four orders of magnitude smaller than the effect of the spin-orbit coupling of the scattering interaction studied in this paper, which is on the scale of tens of MHz.
\bibliographystyle{apsrev4-1}
%\bibliography{biblio.bib}

\begin{thebibliography}{57}%
\makeatletter
\providecommand \@ifxundefined [1]{%
 \@ifx{#1\undefined}
}%
\providecommand \@ifnum [1]{%
 \ifnum #1\expandafter \@firstoftwo
 \else \expandafter \@secondoftwo
 \fi
}%
\providecommand \@ifx [1]{%
 \ifx #1\expandafter \@firstoftwo
 \else \expandafter \@secondoftwo
 \fi
}%
\providecommand \natexlab [1]{#1}%
\providecommand \enquote  [1]{``#1''}%
\providecommand \bibnamefont  [1]{#1}%
\providecommand \bibfnamefont [1]{#1}%
\providecommand \citenamefont [1]{#1}%
\providecommand \href@noop [0]{\@secondoftwo}%
\providecommand \href [0]{\begingroup \@sanitize@url \@href}%
\providecommand \@href[1]{\@@startlink{#1}\@@href}%
\providecommand \@@href[1]{\endgroup#1\@@endlink}%
\providecommand \@sanitize@url [0]{\catcode `\\12\catcode `\$12\catcode
  `\&12\catcode `\#12\catcode `\^12\catcode `\_12\catcode `\%12\relax}%
\providecommand \@@startlink[1]{}%
\providecommand \@@endlink[0]{}%
\providecommand \url  [0]{\begingroup\@sanitize@url \@url }%
\providecommand \@url [1]{\endgroup\@href {#1}{\urlprefix }}%
\providecommand \urlprefix  [0]{URL }%
\providecommand \Eprint [0]{\href }%
\providecommand \doibase [0]{http://dx.doi.org/}%
\providecommand \selectlanguage [0]{\@gobble}%
\providecommand \bibinfo  [0]{\@secondoftwo}%
\providecommand \bibfield  [0]{\@secondoftwo}%
\providecommand \translation [1]{[#1]}%
\providecommand \BibitemOpen [0]{}%
\providecommand \bibitemStop [0]{}%
\providecommand \bibitemNoStop [0]{.\EOS\space}%
\providecommand \EOS [0]{\spacefactor3000\relax}%
\providecommand \BibitemShut  [1]{\csname bibitem#1\endcsname}%
\let\auto@bib@innerbib\@empty
%</preamble>
\bibitem [{\citenamefont {Greene}\ \emph {et~al.}(2000)\citenamefont {Greene},
  \citenamefont {Dickinson},\ and\ \citenamefont
  {Sadeghpour}}]{greene_creation_2000}%
  \BibitemOpen
  \bibfield  {author} {\bibinfo {author} {\bibfnamefont {C.~H.}\ \bibnamefont
  {Greene}}, \bibinfo {author} {\bibfnamefont {A.~S.}\ \bibnamefont
  {Dickinson}}, \ and\ \bibinfo {author} {\bibfnamefont {H.~R.}\ \bibnamefont
  {Sadeghpour}},\ }\href {https://link.aps.org/doi/10.1103/PhysRevLett.85.2458}
  {\bibfield  {journal} {\bibinfo  {journal} {Phys. Rev. Lett.}\ }\textbf
  {\bibinfo {volume} {85}},\ \bibinfo {pages} {2458} (\bibinfo {year}
  {2000})}\BibitemShut {NoStop}%
\bibitem [{\citenamefont {Hamilton}\ \emph {et~al.}(2002)\citenamefont
  {Hamilton}, \citenamefont {Greene},\ and\ \citenamefont
  {Sadeghpour}}]{hamilton_shape-resonance-induced_2002}%
  \BibitemOpen
  \bibfield  {author} {\bibinfo {author} {\bibfnamefont {E.~L.}\ \bibnamefont
  {Hamilton}}, \bibinfo {author} {\bibfnamefont {C.~H.}\ \bibnamefont
  {Greene}}, \ and\ \bibinfo {author} {\bibfnamefont {H.~R.}\ \bibnamefont
  {Sadeghpour}},\ }\href
  {iopscience.iop.org/article/10.1088/0953-4075/35/10/102} {\bibfield
  {journal} {\bibinfo  {journal} {J. Phys. B}\ }\textbf {\bibinfo {volume}
  {35}},\ \bibinfo {pages} {L199} (\bibinfo {year} {2002})}\BibitemShut
  {NoStop}%
\bibitem [{\citenamefont {Chibisov}\ \emph {et~al.}(2002)\citenamefont
  {Chibisov}, \citenamefont {Khuskivadze},\ and\ \citenamefont
  {Fabrikant}}]{Chibisov2002}%
  \BibitemOpen
  \bibfield  {author} {\bibinfo {author} {\bibfnamefont {M.~I.}\ \bibnamefont
  {Chibisov}}, \bibinfo {author} {\bibfnamefont {A.~A.}\ \bibnamefont
  {Khuskivadze}}, \ and\ \bibinfo {author} {\bibfnamefont {I.~I.}\ \bibnamefont
  {Fabrikant}},\ }\href
  {http://iopscience.iop.org/article/10.1088/0953-4075/35/10/101/meta}
  {\bibfield  {journal} {\bibinfo  {journal} {J. Phys. B}\ }\textbf {\bibinfo
  {volume} {35}},\ \bibinfo {pages} {L193} (\bibinfo {year}
  {2002})}\BibitemShut {NoStop}%
\bibitem [{\citenamefont {Bendkowsky}\ \emph {et~al.}(2009)\citenamefont
  {Bendkowsky}, \citenamefont {Butscher}, \citenamefont {Nipper}, \citenamefont
  {Shaffer}, \citenamefont {L{\"{o}}w},\ and\ \citenamefont
  {Pfau}}]{Bendkowsky2009}%
  \BibitemOpen
  \bibfield  {author} {\bibinfo {author} {\bibfnamefont {V.}~\bibnamefont
  {Bendkowsky}}, \bibinfo {author} {\bibfnamefont {B.}~\bibnamefont
  {Butscher}}, \bibinfo {author} {\bibfnamefont {J.}~\bibnamefont {Nipper}},
  \bibinfo {author} {\bibfnamefont {J.~P.}\ \bibnamefont {Shaffer}}, \bibinfo
  {author} {\bibfnamefont {R.}~\bibnamefont {L{\"{o}}w}}, \ and\ \bibinfo
  {author} {\bibfnamefont {T.}~\bibnamefont {Pfau}},\ }\href
  {https://www.nature.com/articles/nature07945} {\bibfield  {journal} {\bibinfo
   {journal} {Nature (London)}\ }\textbf {\bibinfo {volume} {458}},\ \bibinfo
  {pages} {1005} (\bibinfo {year} {2009})}\BibitemShut {NoStop}%
\bibitem [{\citenamefont {Tallant}\ \emph {et~al.}(2012)\citenamefont
  {Tallant}, \citenamefont {Rittenhouse}, \citenamefont {Booth}, \citenamefont
  {Sadeghpour},\ and\ \citenamefont {Shaffer}}]{Tallant2012}%
  \BibitemOpen
  \bibfield  {author} {\bibinfo {author} {\bibfnamefont {J.}~\bibnamefont
  {Tallant}}, \bibinfo {author} {\bibfnamefont {S.~T.}\ \bibnamefont
  {Rittenhouse}}, \bibinfo {author} {\bibfnamefont {D.}~\bibnamefont {Booth}},
  \bibinfo {author} {\bibfnamefont {H.~R.}\ \bibnamefont {Sadeghpour}}, \ and\
  \bibinfo {author} {\bibfnamefont {J.~P.}\ \bibnamefont {Shaffer}},\ }\href
  {https://link.aps.org/pdf/10.1103/PhysRevLett.109.173202} {\bibfield
  {journal} {\bibinfo  {journal} {Phys. Rev. Lett.}\ }\textbf {\bibinfo
  {volume} {109}},\ \bibinfo {pages} {173202} (\bibinfo {year}
  {2012})}\BibitemShut {NoStop}%
\bibitem [{\citenamefont {DeSalvo}\ \emph {et~al.}(2015)\citenamefont
  {DeSalvo}, \citenamefont {Aman}, \citenamefont {Dunning}, \citenamefont
  {Killian}, \citenamefont {Sadeghpour}, \citenamefont {Yoshida},\ and\
  \citenamefont {Burgd{\"{o}}rfer}}]{DeSalvo2015}%
  \BibitemOpen
  \bibfield  {author} {\bibinfo {author} {\bibfnamefont {B.~J.}\ \bibnamefont
  {DeSalvo}}, \bibinfo {author} {\bibfnamefont {J.~A.}\ \bibnamefont {Aman}},
  \bibinfo {author} {\bibfnamefont {F.~B.}\ \bibnamefont {Dunning}}, \bibinfo
  {author} {\bibfnamefont {T.~C.}\ \bibnamefont {Killian}}, \bibinfo {author}
  {\bibfnamefont {H.~R.}\ \bibnamefont {Sadeghpour}}, \bibinfo {author}
  {\bibfnamefont {S.}~\bibnamefont {Yoshida}}, \ and\ \bibinfo {author}
  {\bibfnamefont {J.}~\bibnamefont {Burgd{\"{o}}rfer}},\ }\href
  {https://link.aps.org/doi/10.1103/PhysRevA.92.031403} {\bibfield  {journal}
  {\bibinfo  {journal} {Phys. Rev. A}\ }\textbf {\bibinfo {volume} {92}},\
  \bibinfo {pages} {031403(R)} (\bibinfo {year} {2015})}\BibitemShut {NoStop}%
\bibitem [{\citenamefont {Butscher}\ \emph {et~al.}(2010)\citenamefont
  {Butscher}, \citenamefont {Nipper}, \citenamefont {Balewski}, \citenamefont
  {Kukota}, \citenamefont {Bendkowsky}, \citenamefont {L{\"{o}}w},\ and\
  \citenamefont {Pfau}}]{Butscher2010}%
  \BibitemOpen
  \bibfield  {author} {\bibinfo {author} {\bibfnamefont {B.}~\bibnamefont
  {Butscher}}, \bibinfo {author} {\bibfnamefont {J.}~\bibnamefont {Nipper}},
  \bibinfo {author} {\bibfnamefont {J.~B.}\ \bibnamefont {Balewski}}, \bibinfo
  {author} {\bibfnamefont {L.}~\bibnamefont {Kukota}}, \bibinfo {author}
  {\bibfnamefont {V.}~\bibnamefont {Bendkowsky}}, \bibinfo {author}
  {\bibfnamefont {R.}~\bibnamefont {L{\"{o}}w}}, \ and\ \bibinfo {author}
  {\bibfnamefont {T.}~\bibnamefont {Pfau}},\ }\href
  {https://www.nature.com/articles/nphys1828} {\bibfield  {journal} {\bibinfo
  {journal} {Nature Physics}\ }\textbf {\bibinfo {volume} {6}},\ \bibinfo
  {pages} {970} (\bibinfo {year} {2010})}\BibitemShut {NoStop}%
\bibitem [{\citenamefont {Butscher}\ \emph {et~al.}(2011)\citenamefont
  {Butscher}, \citenamefont {Bendkowsky}, \citenamefont {Nipper}, \citenamefont
  {Balewski}, \citenamefont {Kukota}, \citenamefont {L{\"{o}}w}, \citenamefont
  {Pfau}, \citenamefont {Li}, \citenamefont {Pohl},\ and\ \citenamefont
  {Rost}}]{Butscher2011}%
  \BibitemOpen
  \bibfield  {author} {\bibinfo {author} {\bibfnamefont {B.}~\bibnamefont
  {Butscher}}, \bibinfo {author} {\bibfnamefont {V.}~\bibnamefont
  {Bendkowsky}}, \bibinfo {author} {\bibfnamefont {J.}~\bibnamefont {Nipper}},
  \bibinfo {author} {\bibfnamefont {J.~B.}\ \bibnamefont {Balewski}}, \bibinfo
  {author} {\bibfnamefont {L.}~\bibnamefont {Kukota}}, \bibinfo {author}
  {\bibfnamefont {R.}~\bibnamefont {L{\"{o}}w}}, \bibinfo {author}
  {\bibfnamefont {T.}~\bibnamefont {Pfau}}, \bibinfo {author} {\bibfnamefont
  {W.}~\bibnamefont {Li}}, \bibinfo {author} {\bibfnamefont {T.}~\bibnamefont
  {Pohl}}, \ and\ \bibinfo {author} {\bibfnamefont {J.~M.}\ \bibnamefont
  {Rost}},\ }\href {iopscience.iop.org/article/10.1088/0953-4075/44/18/184004}
  {\bibfield  {journal} {\bibinfo  {journal} {J. Phys. B}\ }\textbf {\bibinfo
  {volume} {44}},\ \bibinfo {pages} {184004} (\bibinfo {year}
  {2011})}\BibitemShut {NoStop}%
\bibitem [{\citenamefont {Camargo}\ \emph {et~al.}(2016)\citenamefont
  {Camargo}, \citenamefont {Whalen}, \citenamefont {Ding}, \citenamefont
  {Sadeghpour}, \citenamefont {Yoshida}, \citenamefont {Burgd{\"{o}}rfer},
  \citenamefont {Dunning},\ and\ \citenamefont {Killian}}]{Camargo2016}%
  \BibitemOpen
  \bibfield  {author} {\bibinfo {author} {\bibfnamefont {F.}~\bibnamefont
  {Camargo}}, \bibinfo {author} {\bibfnamefont {J.~D.}\ \bibnamefont {Whalen}},
  \bibinfo {author} {\bibfnamefont {R.}~\bibnamefont {Ding}}, \bibinfo {author}
  {\bibfnamefont {H.~R.}\ \bibnamefont {Sadeghpour}}, \bibinfo {author}
  {\bibfnamefont {S.}~\bibnamefont {Yoshida}}, \bibinfo {author} {\bibfnamefont
  {J.}~\bibnamefont {Burgd{\"{o}}rfer}}, \bibinfo {author} {\bibfnamefont
  {F.~B.}\ \bibnamefont {Dunning}}, \ and\ \bibinfo {author} {\bibfnamefont
  {T.~C.}\ \bibnamefont {Killian}},\ }\href
  {https://journals.aps.org/pra/abstract/10.1103/PhysRevA.93.022702} {\bibfield
   {journal} {\bibinfo  {journal} {Phys. Rev. A}\ }\textbf {\bibinfo {volume}
  {93}},\ \bibinfo {pages} {022702} (\bibinfo {year} {2016})}\BibitemShut
  {NoStop}%
\bibitem [{\citenamefont {Whalen}\ \emph {et~al.}(2017)\citenamefont {Whalen},
  \citenamefont {Camargo}, \citenamefont {Ding}, \citenamefont {Killian},
  \citenamefont {Dunning}, \citenamefont {P{\'{e}}rez-R{\'{i}}os},
  \citenamefont {Yoshida},\ and\ \citenamefont
  {Burgd{\"{o}}rfer}}]{Whalen2017}%
  \BibitemOpen
  \bibfield  {author} {\bibinfo {author} {\bibfnamefont {J.~D.}\ \bibnamefont
  {Whalen}}, \bibinfo {author} {\bibfnamefont {F.}~\bibnamefont {Camargo}},
  \bibinfo {author} {\bibfnamefont {R.}~\bibnamefont {Ding}}, \bibinfo {author}
  {\bibfnamefont {T.~C.}\ \bibnamefont {Killian}}, \bibinfo {author}
  {\bibfnamefont {F.~B.}\ \bibnamefont {Dunning}}, \bibinfo {author}
  {\bibfnamefont {J.}~\bibnamefont {P{\'{e}}rez-R{\'{i}}os}}, \bibinfo {author}
  {\bibfnamefont {S.}~\bibnamefont {Yoshida}}, \ and\ \bibinfo {author}
  {\bibfnamefont {J.}~\bibnamefont {Burgd{\"{o}}rfer}},\ }\href
  {https://link.aps.org/doi/10.1103/PhysRevA.96.042702} {\bibfield  {journal}
  {\bibinfo  {journal} {Phys. Rev. A}\ }\textbf {\bibinfo {volume} {96}},\
  \bibinfo {pages} {042702} (\bibinfo {year} {2017})}\BibitemShut {NoStop}%
\bibitem [{\citenamefont {Li}\ \emph {et~al.}(2011)\citenamefont {Li},
  \citenamefont {Pohl}, \citenamefont {Rost}, \citenamefont {Rittenhouse},
  \citenamefont {Sadeghpour}, \citenamefont {Nipper}, \citenamefont {Butscher},
  \citenamefont {Balewski}, \citenamefont {Bendkowsky}, \citenamefont
  {L{\"{o}}w},\ and\ \citenamefont {Pfau}}]{li_homonuclear_2011}%
  \BibitemOpen
  \bibfield  {author} {\bibinfo {author} {\bibfnamefont {W.}~\bibnamefont
  {Li}}, \bibinfo {author} {\bibfnamefont {T.}~\bibnamefont {Pohl}}, \bibinfo
  {author} {\bibfnamefont {J.~M.}\ \bibnamefont {Rost}}, \bibinfo {author}
  {\bibfnamefont {S.~T.}\ \bibnamefont {Rittenhouse}}, \bibinfo {author}
  {\bibfnamefont {H.~R.}\ \bibnamefont {Sadeghpour}}, \bibinfo {author}
  {\bibfnamefont {J.}~\bibnamefont {Nipper}}, \bibinfo {author} {\bibfnamefont
  {B.}~\bibnamefont {Butscher}}, \bibinfo {author} {\bibfnamefont {J.~B.}\
  \bibnamefont {Balewski}}, \bibinfo {author} {\bibfnamefont {V.}~\bibnamefont
  {Bendkowsky}}, \bibinfo {author} {\bibfnamefont {R.}~\bibnamefont
  {L{\"{o}}w}}, \ and\ \bibinfo {author} {\bibfnamefont {T.}~\bibnamefont
  {Pfau}},\ }\href {science.sciencemag.org/content/334/6059/1110} {\bibfield
  {journal} {\bibinfo  {journal} {Science}\ }\textbf {\bibinfo {volume}
  {334}},\ \bibinfo {pages} {1110} (\bibinfo {year} {2011})}\BibitemShut
  {NoStop}%
\bibitem [{\citenamefont {Booth}\ \emph {et~al.}(2015)\citenamefont {Booth},
  \citenamefont {Rittenhouse}, \citenamefont {Yang}, \citenamefont
  {Sadeghpour},\ and\ \citenamefont {Shaffer}}]{Booth2015}%
  \BibitemOpen
  \bibfield  {author} {\bibinfo {author} {\bibfnamefont {D.}~\bibnamefont
  {Booth}}, \bibinfo {author} {\bibfnamefont {S.~T.}\ \bibnamefont
  {Rittenhouse}}, \bibinfo {author} {\bibfnamefont {J.}~\bibnamefont {Yang}},
  \bibinfo {author} {\bibfnamefont {H.~R.}\ \bibnamefont {Sadeghpour}}, \ and\
  \bibinfo {author} {\bibfnamefont {J.~P.}\ \bibnamefont {Shaffer}},\ }\href
  {http://science.sciencemag.org/content/348/6230/99} {\bibfield  {journal}
  {\bibinfo  {journal} {Science}\ }\textbf {\bibinfo {volume} {348}},\ \bibinfo
  {pages} {99} (\bibinfo {year} {2015})}\BibitemShut {NoStop}%
\bibitem [{\citenamefont {Anderson}\ \emph {et~al.}(2014)\citenamefont
  {Anderson}, \citenamefont {Miller},\ and\ \citenamefont
  {Raithel}}]{Anderson2014a}%
  \BibitemOpen
  \bibfield  {author} {\bibinfo {author} {\bibfnamefont {D.~A.}\ \bibnamefont
  {Anderson}}, \bibinfo {author} {\bibfnamefont {S.~A.}\ \bibnamefont
  {Miller}}, \ and\ \bibinfo {author} {\bibfnamefont {G.}~\bibnamefont
  {Raithel}},\ }\href
  {https://journals.aps.org/prl/abstract/10.1103/PhysRevLett.112.163201}
  {\bibfield  {journal} {\bibinfo  {journal} {Phys. Rev. Lett.}\ }\textbf
  {\bibinfo {volume} {112}},\ \bibinfo {pages} {163201} (\bibinfo {year}
  {2014})}\BibitemShut {NoStop}%
\bibitem [{\citenamefont {Bendkowsky}\ \emph {et~al.}(2010)\citenamefont
  {Bendkowsky}, \citenamefont {Butscher}, \citenamefont {Nipper}, \citenamefont
  {Balewski}, \citenamefont {Shaffer}, \citenamefont {L{\"{o}}w}, \citenamefont
  {Pfau}, \citenamefont {Li}, \citenamefont {Stanojevic}, \citenamefont
  {Pohl},\ and\ \citenamefont {Rost}}]{Bendkowsky2010}%
  \BibitemOpen
  \bibfield  {author} {\bibinfo {author} {\bibfnamefont {V.}~\bibnamefont
  {Bendkowsky}}, \bibinfo {author} {\bibfnamefont {B.}~\bibnamefont
  {Butscher}}, \bibinfo {author} {\bibfnamefont {J.}~\bibnamefont {Nipper}},
  \bibinfo {author} {\bibfnamefont {J.~B.}\ \bibnamefont {Balewski}}, \bibinfo
  {author} {\bibfnamefont {J.~P.}\ \bibnamefont {Shaffer}}, \bibinfo {author}
  {\bibfnamefont {R.}~\bibnamefont {L{\"{o}}w}}, \bibinfo {author}
  {\bibfnamefont {T.}~\bibnamefont {Pfau}}, \bibinfo {author} {\bibfnamefont
  {W.}~\bibnamefont {Li}}, \bibinfo {author} {\bibfnamefont {J.}~\bibnamefont
  {Stanojevic}}, \bibinfo {author} {\bibfnamefont {T.}~\bibnamefont {Pohl}}, \
  and\ \bibinfo {author} {\bibfnamefont {J.~M.}\ \bibnamefont {Rost}},\ }\href
  {https://link.aps.org/doi/10.1103/PhysRevLett.105.163201} {\bibfield
  {journal} {\bibinfo  {journal} {Phys. Rev. Lett.}\ }\textbf {\bibinfo
  {volume} {105}},\ \bibinfo {pages} {163201} (\bibinfo {year}
  {2010})}\BibitemShut {NoStop}%
\bibitem [{\citenamefont {Gaj}\ \emph {et~al.}(2014)\citenamefont {Gaj},
  \citenamefont {Krupp}, \citenamefont {Balewski}, \citenamefont {L{\"{o}}w},
  \citenamefont {Hofferberth},\ and\ \citenamefont
  {Pfau}}]{gaj_molecular_2014}%
  \BibitemOpen
  \bibfield  {author} {\bibinfo {author} {\bibfnamefont {A.}~\bibnamefont
  {Gaj}}, \bibinfo {author} {\bibfnamefont {A.~T.}\ \bibnamefont {Krupp}},
  \bibinfo {author} {\bibfnamefont {J.~B.}\ \bibnamefont {Balewski}}, \bibinfo
  {author} {\bibfnamefont {R.}~\bibnamefont {L{\"{o}}w}}, \bibinfo {author}
  {\bibfnamefont {S.}~\bibnamefont {Hofferberth}}, \ and\ \bibinfo {author}
  {\bibfnamefont {T.}~\bibnamefont {Pfau}},\ }\href
  {https://www.nature.com/articles/ncomms5546} {\bibfield  {journal} {\bibinfo
  {journal} {Nat. Commun.}\ }\textbf {\bibinfo {volume} {5}},\ \bibinfo {pages}
  {5546} (\bibinfo {year} {2014})}\BibitemShut {NoStop}%
\bibitem [{\citenamefont {Krupp}\ \emph {et~al.}(2014)\citenamefont {Krupp},
  \citenamefont {Gaj}, \citenamefont {Balewski}, \citenamefont
  {Ilzh{\"{o}}fer}, \citenamefont {Hofferberth}, \citenamefont {L{\"{o}}w},
  \citenamefont {Pfau}, \citenamefont {Kurz},\ and\ \citenamefont
  {Schmelcher}}]{krupp_alignment_2014}%
  \BibitemOpen
  \bibfield  {author} {\bibinfo {author} {\bibfnamefont {A.~T.}\ \bibnamefont
  {Krupp}}, \bibinfo {author} {\bibfnamefont {A.}~\bibnamefont {Gaj}}, \bibinfo
  {author} {\bibfnamefont {J.~B.}\ \bibnamefont {Balewski}}, \bibinfo {author}
  {\bibfnamefont {P.}~\bibnamefont {Ilzh{\"{o}}fer}}, \bibinfo {author}
  {\bibfnamefont {S.}~\bibnamefont {Hofferberth}}, \bibinfo {author}
  {\bibfnamefont {R.}~\bibnamefont {L{\"{o}}w}}, \bibinfo {author}
  {\bibfnamefont {T.}~\bibnamefont {Pfau}}, \bibinfo {author} {\bibfnamefont
  {M.}~\bibnamefont {Kurz}}, \ and\ \bibinfo {author} {\bibfnamefont
  {P.}~\bibnamefont {Schmelcher}},\ }\href
  {https://link.aps.org/doi/10.1103/PhysRevLett.112.143008} {\bibfield
  {journal} {\bibinfo  {journal} {Phys. Rev. Lett.}\ }\textbf {\bibinfo
  {volume} {112}},\ \bibinfo {pages} {143008} (\bibinfo {year}
  {2014})}\BibitemShut {NoStop}%
\bibitem [{\citenamefont {Gaj}\ \emph {et~al.}(2015)\citenamefont {Gaj},
  \citenamefont {Krupp}, \citenamefont {Ilzh{\"{o}}fer}, \citenamefont
  {L{\"{o}}w}, \citenamefont {Hofferberth},\ and\ \citenamefont
  {Pfau}}]{Gaj2015}%
  \BibitemOpen
  \bibfield  {author} {\bibinfo {author} {\bibfnamefont {A.}~\bibnamefont
  {Gaj}}, \bibinfo {author} {\bibfnamefont {A.~T.}\ \bibnamefont {Krupp}},
  \bibinfo {author} {\bibfnamefont {P.}~\bibnamefont {Ilzh{\"{o}}fer}},
  \bibinfo {author} {\bibfnamefont {R.}~\bibnamefont {L{\"{o}}w}}, \bibinfo
  {author} {\bibfnamefont {S.}~\bibnamefont {Hofferberth}}, \ and\ \bibinfo
  {author} {\bibfnamefont {T.}~\bibnamefont {Pfau}},\ }\href
  {https://journals.aps.org/prl/abstract/10.1103/PhysRevLett.115.023001}
  {\bibfield  {journal} {\bibinfo  {journal} {Phys. Rev. Lett.}\ }\textbf
  {\bibinfo {volume} {115}},\ \bibinfo {pages} {023001} (\bibinfo {year}
  {2015})}\BibitemShut {NoStop}%
\bibitem [{\citenamefont {Niederpr{\"{u}}m}\ \emph
  {et~al.}(2016{\natexlab{a}})\citenamefont {Niederpr{\"{u}}m}, \citenamefont
  {Thomas}, \citenamefont {Eichert}, \citenamefont {Lippe}, \citenamefont
  {P{\'{e}}rez-R{\'{i}}os}, \citenamefont {Greene},\ and\ \citenamefont
  {Ott}}]{niederprum_observation_2016}%
  \BibitemOpen
  \bibfield  {author} {\bibinfo {author} {\bibfnamefont {T.}~\bibnamefont
  {Niederpr{\"{u}}m}}, \bibinfo {author} {\bibfnamefont {O.}~\bibnamefont
  {Thomas}}, \bibinfo {author} {\bibfnamefont {T.}~\bibnamefont {Eichert}},
  \bibinfo {author} {\bibfnamefont {C.}~\bibnamefont {Lippe}}, \bibinfo
  {author} {\bibfnamefont {J.}~\bibnamefont {P{\'{e}}rez-R{\'{i}}os}}, \bibinfo
  {author} {\bibfnamefont {C.~H.}\ \bibnamefont {Greene}}, \ and\ \bibinfo
  {author} {\bibfnamefont {H.}~\bibnamefont {Ott}},\ }\href
  {https://www.nature.com/articles/ncomms12820} {\bibfield  {journal} {\bibinfo
   {journal} {Nat. Commun.}\ }\textbf {\bibinfo {volume} {7}},\ \bibinfo
  {pages} {12820} (\bibinfo {year} {2016}{\natexlab{a}})}\BibitemShut {NoStop}%
\bibitem [{\citenamefont {Manthey}\ \emph {et~al.}(2015)\citenamefont
  {Manthey}, \citenamefont {Niederpr{\"{u}}m}, \citenamefont {Thomas},\ and\
  \citenamefont {Ott}}]{Manthey2015}%
  \BibitemOpen
  \bibfield  {author} {\bibinfo {author} {\bibfnamefont {T.}~\bibnamefont
  {Manthey}}, \bibinfo {author} {\bibfnamefont {T.}~\bibnamefont
  {Niederpr{\"{u}}m}}, \bibinfo {author} {\bibfnamefont {O.}~\bibnamefont
  {Thomas}}, \ and\ \bibinfo {author} {\bibfnamefont {H.}~\bibnamefont {Ott}},\
  }\href {iopscience.iop.org/article/10.1088/1367-2630/17/10/103024} {\bibfield
   {journal} {\bibinfo  {journal} {New J. Phys.}\ }\textbf {\bibinfo {volume}
  {17}},\ \bibinfo {pages} {103024} (\bibinfo {year} {2015})}\BibitemShut
  {NoStop}%
\bibitem [{\citenamefont {Niederpr{\"{u}}m}\ \emph
  {et~al.}(2016{\natexlab{b}})\citenamefont {Niederpr{\"{u}}m}, \citenamefont
  {Thomas}, \citenamefont {Eichert},\ and\ \citenamefont
  {Ott}}]{niederprum_rydberg_2016}%
  \BibitemOpen
  \bibfield  {author} {\bibinfo {author} {\bibfnamefont {T.}~\bibnamefont
  {Niederpr{\"{u}}m}}, \bibinfo {author} {\bibfnamefont {O.}~\bibnamefont
  {Thomas}}, \bibinfo {author} {\bibfnamefont {T.}~\bibnamefont {Eichert}}, \
  and\ \bibinfo {author} {\bibfnamefont {H.}~\bibnamefont {Ott}},\ }\href
  {https://link.aps.org/doi/10.1103/PhysRevLett.117.123002} {\bibfield
  {journal} {\bibinfo  {journal} {Phys. Rev. Lett.}\ }\textbf {\bibinfo
  {volume} {117}},\ \bibinfo {pages} {123002} (\bibinfo {year}
  {2016}{\natexlab{b}})}\BibitemShut {NoStop}%
\bibitem [{\citenamefont {Schlagm{\"{u}}ller}\ \emph
  {et~al.}(2016{\natexlab{a}})\citenamefont {Schlagm{\"{u}}ller}, \citenamefont
  {Liebisch}, \citenamefont {Engel}, \citenamefont {Kleinbach}, \citenamefont
  {B{\"{o}}ttcher}, \citenamefont {Hermann}, \citenamefont {Westphal},
  \citenamefont {Gaj}, \citenamefont {L{\"{o}}w}, \citenamefont {Hofferberth},
  \citenamefont {Pfau}, \citenamefont {P{\'{e}}rez-R{\'{i}}os},\ and\
  \citenamefont {Greene}}]{Schlagmuller2016x}%
  \BibitemOpen
  \bibfield  {author} {\bibinfo {author} {\bibfnamefont {M.}~\bibnamefont
  {Schlagm{\"{u}}ller}}, \bibinfo {author} {\bibfnamefont {T.~C.}\ \bibnamefont
  {Liebisch}}, \bibinfo {author} {\bibfnamefont {F.}~\bibnamefont {Engel}},
  \bibinfo {author} {\bibfnamefont {K.~S.}\ \bibnamefont {Kleinbach}}, \bibinfo
  {author} {\bibfnamefont {F.}~\bibnamefont {B{\"{o}}ttcher}}, \bibinfo
  {author} {\bibfnamefont {U.}~\bibnamefont {Hermann}}, \bibinfo {author}
  {\bibfnamefont {K.~M.}\ \bibnamefont {Westphal}}, \bibinfo {author}
  {\bibfnamefont {A.}~\bibnamefont {Gaj}}, \bibinfo {author} {\bibfnamefont
  {R.}~\bibnamefont {L{\"{o}}w}}, \bibinfo {author} {\bibfnamefont
  {S.}~\bibnamefont {Hofferberth}}, \bibinfo {author} {\bibfnamefont
  {T.}~\bibnamefont {Pfau}}, \bibinfo {author} {\bibfnamefont {J.}~\bibnamefont
  {P{\'{e}}rez-R{\'{i}}os}}, \ and\ \bibinfo {author} {\bibfnamefont {C.~H.}\
  \bibnamefont {Greene}},\ }\href
  {https://link.aps.org/doi/10.1103/PhysRevX.6.031020} {\bibfield  {journal}
  {\bibinfo  {journal} {Phys. Rev. X}\ }\textbf {\bibinfo {volume} {6}},\
  \bibinfo {pages} {031020} (\bibinfo {year} {2016}{\natexlab{a}})}\BibitemShut
  {NoStop}%
\bibitem [{\citenamefont {Mirgorodskiy}\ \emph {et~al.}(2017)\citenamefont
  {Mirgorodskiy}, \citenamefont {Christaller}, \citenamefont {Braun},
  \citenamefont {Paris-Mandoki}, \citenamefont {Tresp},\ and\ \citenamefont
  {Hofferberth}}]{Mirgorodskiy2017}%
  \BibitemOpen
  \bibfield  {author} {\bibinfo {author} {\bibfnamefont {I.}~\bibnamefont
  {Mirgorodskiy}}, \bibinfo {author} {\bibfnamefont {F.}~\bibnamefont
  {Christaller}}, \bibinfo {author} {\bibfnamefont {C.}~\bibnamefont {Braun}},
  \bibinfo {author} {\bibfnamefont {A.}~\bibnamefont {Paris-Mandoki}}, \bibinfo
  {author} {\bibfnamefont {C.}~\bibnamefont {Tresp}}, \ and\ \bibinfo {author}
  {\bibfnamefont {S.}~\bibnamefont {Hofferberth}},\ }\href
  {link.aps.org/pdf/10.1103/PhysRevA.96.011402} {\bibfield  {journal} {\bibinfo
   {journal} {Phys. Rev. A}\ }\textbf {\bibinfo {volume} {96}},\ \bibinfo
  {pages} {011402(R)} (\bibinfo {year} {2017})}\BibitemShut {NoStop}%
\bibitem [{\citenamefont {Balewski}\ \emph {et~al.}(2013)\citenamefont
  {Balewski}, \citenamefont {Krupp}, \citenamefont {Gaj}, \citenamefont
  {Peter}, \citenamefont {B{\"{u}}chler}, \citenamefont {L{\"{o}}w},
  \citenamefont {Hofferberth},\ and\ \citenamefont
  {Pfau}}]{balewski_coupling_2013}%
  \BibitemOpen
  \bibfield  {author} {\bibinfo {author} {\bibfnamefont {J.~B.}\ \bibnamefont
  {Balewski}}, \bibinfo {author} {\bibfnamefont {A.~T.}\ \bibnamefont {Krupp}},
  \bibinfo {author} {\bibfnamefont {A.}~\bibnamefont {Gaj}}, \bibinfo {author}
  {\bibfnamefont {D.}~\bibnamefont {Peter}}, \bibinfo {author} {\bibfnamefont
  {H.~P.}\ \bibnamefont {B{\"{u}}chler}}, \bibinfo {author} {\bibfnamefont
  {R.}~\bibnamefont {L{\"{o}}w}}, \bibinfo {author} {\bibfnamefont
  {S.}~\bibnamefont {Hofferberth}}, \ and\ \bibinfo {author} {\bibfnamefont
  {T.}~\bibnamefont {Pfau}},\ }\href
  {https://www.nature.com/articles/nature12592} {\bibfield  {journal} {\bibinfo
   {journal} {Nature (London)}\ }\textbf {\bibinfo {volume} {502}},\ \bibinfo
  {pages} {664} (\bibinfo {year} {2013})}\BibitemShut {NoStop}%
\bibitem [{\citenamefont {Wang}\ \emph {et~al.}(2015)\citenamefont {Wang},
  \citenamefont {Gacesa},\ and\ \citenamefont {C{\^{o}}t{\'{e}}}}]{Wang2015}%
  \BibitemOpen
  \bibfield  {author} {\bibinfo {author} {\bibfnamefont {J.}~\bibnamefont
  {Wang}}, \bibinfo {author} {\bibfnamefont {M.}~\bibnamefont {Gacesa}}, \ and\
  \bibinfo {author} {\bibfnamefont {R.}~\bibnamefont {C{\^{o}}t{\'{e}}}},\
  }\href {link.aps.org/pdf/10.1103/PhysRevLett.114.243003} {\bibfield
  {journal} {\bibinfo  {journal} {Phys. Rev. Lett.}\ }\textbf {\bibinfo
  {volume} {114}},\ \bibinfo {pages} {243003} (\bibinfo {year}
  {2015})}\BibitemShut {NoStop}%
\bibitem [{\citenamefont {Camargo}\ \emph {et~al.}(2018)\citenamefont
  {Camargo}, \citenamefont {Schmidt}, \citenamefont {Whalen}, \citenamefont
  {Ding}, \citenamefont {Woehl}, \citenamefont {Yoshida}, \citenamefont
  {Burgd{\"{o}}rfer}, \citenamefont {Dunning}, \citenamefont {Sadeghpour},
  \citenamefont {Demler},\ and\ \citenamefont {Killian}}]{Camargo2017}%
  \BibitemOpen
  \bibfield  {author} {\bibinfo {author} {\bibfnamefont {F.}~\bibnamefont
  {Camargo}}, \bibinfo {author} {\bibfnamefont {R.}~\bibnamefont {Schmidt}},
  \bibinfo {author} {\bibfnamefont {J.~D.}\ \bibnamefont {Whalen}}, \bibinfo
  {author} {\bibfnamefont {R.}~\bibnamefont {Ding}}, \bibinfo {author}
  {\bibfnamefont {G.}~\bibnamefont {Woehl}}, \bibinfo {author} {\bibfnamefont
  {S.}~\bibnamefont {Yoshida}}, \bibinfo {author} {\bibfnamefont
  {J.}~\bibnamefont {Burgd{\"{o}}rfer}}, \bibinfo {author} {\bibfnamefont
  {F.~B.}\ \bibnamefont {Dunning}}, \bibinfo {author} {\bibfnamefont {H.~R.}\
  \bibnamefont {Sadeghpour}}, \bibinfo {author} {\bibfnamefont
  {E.}~\bibnamefont {Demler}}, \ and\ \bibinfo {author} {\bibfnamefont {T.~C.}\
  \bibnamefont {Killian}},\ }\href
  {https://link.aps.org/doi/10.1103/PhysRevLett.120.083401} {\bibfield
  {journal} {\bibinfo  {journal} {Phys. Rev. Lett.}\ }\textbf {\bibinfo
  {volume} {120}},\ \bibinfo {pages} {083401} (\bibinfo {year}
  {2018})}\BibitemShut {NoStop}%
\bibitem [{\citenamefont {Eiles}\ \emph {et~al.}(2016)\citenamefont {Eiles},
  \citenamefont {P{\'{e}}rez-R{\'{i}}os}, \citenamefont {Robicheaux},\ and\
  \citenamefont {Greene}}]{Eiles2016}%
  \BibitemOpen
  \bibfield  {author} {\bibinfo {author} {\bibfnamefont {M.~T.}\ \bibnamefont
  {Eiles}}, \bibinfo {author} {\bibfnamefont {J.}~\bibnamefont
  {P{\'{e}}rez-R{\'{i}}os}}, \bibinfo {author} {\bibfnamefont {F.}~\bibnamefont
  {Robicheaux}}, \ and\ \bibinfo {author} {\bibfnamefont {C.~H.}\ \bibnamefont
  {Greene}},\ }\href
  {http://iopscience.iop.org/article/10.1088/0953-4075/49/11/114005/meta}
  {\bibfield  {journal} {\bibinfo  {journal} {J. Phys. B}\ }\textbf {\bibinfo
  {volume} {49}},\ \bibinfo {pages} {114005} (\bibinfo {year}
  {2016})}\BibitemShut {NoStop}%
\bibitem [{\citenamefont {Schmidt}\ \emph {et~al.}(2016)\citenamefont
  {Schmidt}, \citenamefont {Sadeghpour},\ and\ \citenamefont
  {Demler}}]{Schmidt2016}%
  \BibitemOpen
  \bibfield  {author} {\bibinfo {author} {\bibfnamefont {R.}~\bibnamefont
  {Schmidt}}, \bibinfo {author} {\bibfnamefont {H.~R.}\ \bibnamefont
  {Sadeghpour}}, \ and\ \bibinfo {author} {\bibfnamefont {E.}~\bibnamefont
  {Demler}},\ }\href {https://link.aps.org/doi/10.1103/PhysRevLett.116.105302}
  {\bibfield  {journal} {\bibinfo  {journal} {Phys. Rev. Lett.}\ }\textbf
  {\bibinfo {volume} {116}},\ \bibinfo {pages} {105302} (\bibinfo {year}
  {2016})}\BibitemShut {NoStop}%
\bibitem [{\citenamefont {Schmidt}\ \emph {et~al.}(2018)\citenamefont
  {Schmidt}, \citenamefont {Whalen}, \citenamefont {Ding}, \citenamefont
  {Camargo}, \citenamefont {Woehl}, \citenamefont {Yoshida}, \citenamefont
  {Burgd{\"{o}}rfer}, \citenamefont {Dunning}, \citenamefont {Demler},
  \citenamefont {Sadeghpour},\ and\ \citenamefont {Killian}}]{Schmidt2018}%
  \BibitemOpen
  \bibfield  {author} {\bibinfo {author} {\bibfnamefont {R.}~\bibnamefont
  {Schmidt}}, \bibinfo {author} {\bibfnamefont {J.~D.}\ \bibnamefont {Whalen}},
  \bibinfo {author} {\bibfnamefont {R.}~\bibnamefont {Ding}}, \bibinfo {author}
  {\bibfnamefont {F.}~\bibnamefont {Camargo}}, \bibinfo {author} {\bibfnamefont
  {G.}~\bibnamefont {Woehl}}, \bibinfo {author} {\bibfnamefont
  {S.}~\bibnamefont {Yoshida}}, \bibinfo {author} {\bibfnamefont
  {J.}~\bibnamefont {Burgd{\"{o}}rfer}}, \bibinfo {author} {\bibfnamefont
  {F.~B.}\ \bibnamefont {Dunning}}, \bibinfo {author} {\bibfnamefont
  {E.}~\bibnamefont {Demler}}, \bibinfo {author} {\bibfnamefont {H.~R.}\
  \bibnamefont {Sadeghpour}}, \ and\ \bibinfo {author} {\bibfnamefont {T.~C.}\
  \bibnamefont {Killian}},\ }\href
  {link.aps.org/pdf/10.1103/PhysRevA.97.022707} {\bibfield  {journal} {\bibinfo
   {journal} {Phys. Rev. A}\ }\textbf {\bibinfo {volume} {97}},\ \bibinfo
  {pages} {022707} (\bibinfo {year} {2018})}\BibitemShut {NoStop}%
\bibitem [{\citenamefont {Kleinbach}\ \emph {et~al.}(2018)\citenamefont
  {Kleinbach}, \citenamefont {Engel}, \citenamefont {Dieterle}, \citenamefont
  {L{\"{o}}w}, \citenamefont {Pfau},\ and\ \citenamefont
  {Meinert}}]{Kleinbach2018}%
  \BibitemOpen
  \bibfield  {author} {\bibinfo {author} {\bibfnamefont {K.~S.}\ \bibnamefont
  {Kleinbach}}, \bibinfo {author} {\bibfnamefont {F.}~\bibnamefont {Engel}},
  \bibinfo {author} {\bibfnamefont {T.}~\bibnamefont {Dieterle}}, \bibinfo
  {author} {\bibfnamefont {R.}~\bibnamefont {L{\"{o}}w}}, \bibinfo {author}
  {\bibfnamefont {T.}~\bibnamefont {Pfau}}, \ and\ \bibinfo {author}
  {\bibfnamefont {F.}~\bibnamefont {Meinert}},\ }\href
  {https://link.aps.org/doi/10.1103/PhysRevLett.120.193401} {\bibfield
  {journal} {\bibinfo  {journal} {Phys. Rev. Lett.}\ }\textbf {\bibinfo
  {volume} {120}},\ \bibinfo {pages} {193401} (\bibinfo {year}
  {2018})}\BibitemShut {NoStop}%
\bibitem [{\citenamefont {Schmid}\ \emph {et~al.}(2018)\citenamefont {Schmid},
  \citenamefont {Veit}, \citenamefont {Zuber}, \citenamefont {L{\"{o}}w},
  \citenamefont {Pfau}, \citenamefont {Tarana},\ and\ \citenamefont
  {Tomza}}]{Schmid2018}%
  \BibitemOpen
  \bibfield  {author} {\bibinfo {author} {\bibfnamefont {T.}~\bibnamefont
  {Schmid}}, \bibinfo {author} {\bibfnamefont {C.}~\bibnamefont {Veit}},
  \bibinfo {author} {\bibfnamefont {N.}~\bibnamefont {Zuber}}, \bibinfo
  {author} {\bibfnamefont {R.}~\bibnamefont {L{\"{o}}w}}, \bibinfo {author}
  {\bibfnamefont {T.}~\bibnamefont {Pfau}}, \bibinfo {author} {\bibfnamefont
  {M.}~\bibnamefont {Tarana}}, \ and\ \bibinfo {author} {\bibfnamefont
  {M.}~\bibnamefont {Tomza}},\ }\href
  {https://link.aps.org/doi/10.1103/PhysRevLett.120.153401} {\bibfield
  {journal} {\bibinfo  {journal} {Phys. Rev. Lett.}\ }\textbf {\bibinfo
  {volume} {120}},\ \bibinfo {pages} {153401} (\bibinfo {year}
  {2018})}\BibitemShut {NoStop}%
\bibitem [{\citenamefont {Engel}\ \emph {et~al.}(2018)\citenamefont {Engel},
  \citenamefont {Dieterle}, \citenamefont {Schmid}, \citenamefont {Tomschitz},
  \citenamefont {Veit}, \citenamefont {Zuber}, \citenamefont {L\"ow},
  \citenamefont {Pfau},\ and\ \citenamefont {Meinert}}]{Engel2018}%
  \BibitemOpen
  \bibfield  {author} {\bibinfo {author} {\bibfnamefont {F.}~\bibnamefont
  {Engel}}, \bibinfo {author} {\bibfnamefont {T.}~\bibnamefont {Dieterle}},
  \bibinfo {author} {\bibfnamefont {T.}~\bibnamefont {Schmid}}, \bibinfo
  {author} {\bibfnamefont {C.}~\bibnamefont {Tomschitz}}, \bibinfo {author}
  {\bibfnamefont {C.}~\bibnamefont {Veit}}, \bibinfo {author} {\bibfnamefont
  {N.}~\bibnamefont {Zuber}}, \bibinfo {author} {\bibfnamefont
  {R.}~\bibnamefont {L\"ow}}, \bibinfo {author} {\bibfnamefont
  {T.}~\bibnamefont {Pfau}}, \ and\ \bibinfo {author} {\bibfnamefont
  {F.}~\bibnamefont {Meinert}},\ }\href {\doibase
  10.1103/PhysRevLett.121.193401} {\bibfield  {journal} {\bibinfo  {journal}
  {Phys. Rev. Lett.}\ }\textbf {\bibinfo {volume} {121}},\ \bibinfo {pages}
  {193401} (\bibinfo {year} {2018})}\BibitemShut {NoStop}%
\bibitem [{\citenamefont {S{\'{a}}ndor}\ \emph {et~al.}(2017)\citenamefont
  {S{\'{a}}ndor}, \citenamefont {Gonz{\'{a}}lez-F{\'{e}}rez}, \citenamefont
  {Julienne},\ and\ \citenamefont {Pupillo}}]{Sandor2017}%
  \BibitemOpen
  \bibfield  {author} {\bibinfo {author} {\bibfnamefont {N.}~\bibnamefont
  {S{\'{a}}ndor}}, \bibinfo {author} {\bibfnamefont {R.}~\bibnamefont
  {Gonz{\'{a}}lez-F{\'{e}}rez}}, \bibinfo {author} {\bibfnamefont {P.~S.}\
  \bibnamefont {Julienne}}, \ and\ \bibinfo {author} {\bibfnamefont
  {G.}~\bibnamefont {Pupillo}},\ }\href
  {https://link.aps.org/doi/10.1103/PhysRevA.96.032719} {\bibfield  {journal}
  {\bibinfo  {journal} {Phys. Rev. A}\ }\textbf {\bibinfo {volume} {96}},\
  \bibinfo {pages} {032719} (\bibinfo {year} {2017})}\BibitemShut {NoStop}%
\bibitem [{\citenamefont {Thomas}\ \emph {et~al.}(2018)\citenamefont {Thomas},
  \citenamefont {Lippe}, \citenamefont {Eichert},\ and\ \citenamefont
  {Ott}}]{Thomas2017}%
  \BibitemOpen
  \bibfield  {author} {\bibinfo {author} {\bibfnamefont {O.}~\bibnamefont
  {Thomas}}, \bibinfo {author} {\bibfnamefont {C.}~\bibnamefont {Lippe}},
  \bibinfo {author} {\bibfnamefont {T.}~\bibnamefont {Eichert}}, \ and\
  \bibinfo {author} {\bibfnamefont {H.}~\bibnamefont {Ott}},\ }\href
  {https://www.nature.com/articles/s41467-018-04684-w} {\bibfield  {journal}
  {\bibinfo  {journal} {Nat. Commun.}\ }\textbf {\bibinfo {volume} {9}},\
  \bibinfo {pages} {2238} (\bibinfo {year} {2018})}\BibitemShut {NoStop}%
\bibitem [{\citenamefont {Fermi}(1934)}]{Fermi1934}%
  \BibitemOpen
  \bibfield  {author} {\bibinfo {author} {\bibfnamefont {E.}~\bibnamefont
  {Fermi}},\ }\href {https://link.springer.com/article/10.1007/BF02959829}
  {\bibfield  {journal} {\bibinfo  {journal} {Nuovo Cimento}\ }\textbf
  {\bibinfo {volume} {11}},\ \bibinfo {pages} {157} (\bibinfo {year}
  {1934})}\BibitemShut {NoStop}%
\bibitem [{\citenamefont {Omont}(1977)}]{omont_theory_1977}%
  \BibitemOpen
  \bibfield  {author} {\bibinfo {author} {\bibfnamefont {A.}~\bibnamefont
  {Omont}},\ }\href
  {https://jphys.journaldephysique.org/articles/jphys/abs/1977/11/jphys_1977__38_11_1343_0/jphys_1977__38_11_1343_0.html}
  {\bibfield  {journal} {\bibinfo  {journal} {J. Phys. (Paris)}\ }\textbf
  {\bibinfo {volume} {38}},\ \bibinfo {pages} {1343} (\bibinfo {year}
  {1977})}\BibitemShut {NoStop}%
\bibitem [{\citenamefont {Sa{\ss}mannshausen}\ \emph
  {et~al.}(2015)\citenamefont {Sa{\ss}mannshausen}, \citenamefont {Merkt},\
  and\ \citenamefont {Deiglmayr}}]{Sassmannshausen2015}%
  \BibitemOpen
  \bibfield  {author} {\bibinfo {author} {\bibfnamefont {H.}~\bibnamefont
  {Sa{\ss}mannshausen}}, \bibinfo {author} {\bibfnamefont {F.}~\bibnamefont
  {Merkt}}, \ and\ \bibinfo {author} {\bibfnamefont {J.}~\bibnamefont
  {Deiglmayr}},\ }\href
  {https://link.aps.org/pdf/10.1103/PhysRevLett.114.133201} {\bibfield
  {journal} {\bibinfo  {journal} {Phys. Rev. Lett.}\ }\textbf {\bibinfo
  {volume} {114}},\ \bibinfo {pages} {133201} (\bibinfo {year}
  {2015})}\BibitemShut {NoStop}%
\bibitem [{\citenamefont {B{\"{o}}ttcher}\ \emph {et~al.}(2016)\citenamefont
  {B{\"{o}}ttcher}, \citenamefont {Gaj}, \citenamefont {Westphal},
  \citenamefont {Schlagm{\"{u}}ller}, \citenamefont {Kleinbach}, \citenamefont
  {L{\"{o}}w}, \citenamefont {Liebisch}, \citenamefont {Pfau},\ and\
  \citenamefont {Hofferberth}}]{Boettcher2016}%
  \BibitemOpen
  \bibfield  {author} {\bibinfo {author} {\bibfnamefont {F.}~\bibnamefont
  {B{\"{o}}ttcher}}, \bibinfo {author} {\bibfnamefont {A.}~\bibnamefont {Gaj}},
  \bibinfo {author} {\bibfnamefont {K.~M.}\ \bibnamefont {Westphal}}, \bibinfo
  {author} {\bibfnamefont {M.}~\bibnamefont {Schlagm{\"{u}}ller}}, \bibinfo
  {author} {\bibfnamefont {K.~S.}\ \bibnamefont {Kleinbach}}, \bibinfo {author}
  {\bibfnamefont {R.}~\bibnamefont {L{\"{o}}w}}, \bibinfo {author}
  {\bibfnamefont {T.~C.}\ \bibnamefont {Liebisch}}, \bibinfo {author}
  {\bibfnamefont {T.}~\bibnamefont {Pfau}}, \ and\ \bibinfo {author}
  {\bibfnamefont {S.}~\bibnamefont {Hofferberth}},\ }\href
  {https://link.aps.org/pdf/10.1103/PhysRevA.93.032512} {\bibfield  {journal}
  {\bibinfo  {journal} {Phys. Rev. A}\ }\textbf {\bibinfo {volume} {93}},\
  \bibinfo {pages} {032512} (\bibinfo {year} {2016})}\BibitemShut {NoStop}%
\bibitem [{\citenamefont {Schlagm{\"{u}}ller}\ \emph
  {et~al.}(2016{\natexlab{b}})\citenamefont {Schlagm{\"{u}}ller}, \citenamefont
  {Liebisch}, \citenamefont {Nguyen}, \citenamefont {Lochead}, \citenamefont
  {Engel}, \citenamefont {B{\"{o}}ttcher}, \citenamefont {Westphal},
  \citenamefont {Kleinbach}, \citenamefont {L{\"{o}}w}, \citenamefont
  {Hofferberth}, \citenamefont {Pfau}, \citenamefont {P{\'{e}}rez-R{\'{i}}os},\
  and\ \citenamefont {Greene}}]{Schlagmuller2016}%
  \BibitemOpen
  \bibfield  {author} {\bibinfo {author} {\bibfnamefont {M.}~\bibnamefont
  {Schlagm{\"{u}}ller}}, \bibinfo {author} {\bibfnamefont {T.~C.}\ \bibnamefont
  {Liebisch}}, \bibinfo {author} {\bibfnamefont {H.}~\bibnamefont {Nguyen}},
  \bibinfo {author} {\bibfnamefont {G.}~\bibnamefont {Lochead}}, \bibinfo
  {author} {\bibfnamefont {F.}~\bibnamefont {Engel}}, \bibinfo {author}
  {\bibfnamefont {F.}~\bibnamefont {B{\"{o}}ttcher}}, \bibinfo {author}
  {\bibfnamefont {K.~M.}\ \bibnamefont {Westphal}}, \bibinfo {author}
  {\bibfnamefont {K.~S.}\ \bibnamefont {Kleinbach}}, \bibinfo {author}
  {\bibfnamefont {R.}~\bibnamefont {L{\"{o}}w}}, \bibinfo {author}
  {\bibfnamefont {S.}~\bibnamefont {Hofferberth}}, \bibinfo {author}
  {\bibfnamefont {T.}~\bibnamefont {Pfau}}, \bibinfo {author} {\bibfnamefont
  {J.}~\bibnamefont {P{\'{e}}rez-R{\'{i}}os}}, \ and\ \bibinfo {author}
  {\bibfnamefont {C.~H.}\ \bibnamefont {Greene}},\ }\href
  {https://link.aps.org/pdf/10.1103/PhysRevLett.116.053001} {\bibfield
  {journal} {\bibinfo  {journal} {Phys. Rev. Lett.}\ }\textbf {\bibinfo
  {volume} {116}},\ \bibinfo {pages} {053001} (\bibinfo {year}
  {2016}{\natexlab{b}})}\BibitemShut {NoStop}%
\bibitem [{\citenamefont {MacLennan}\ \emph {et~al.}(shed)\citenamefont
  {MacLennan}, \citenamefont {Chen},\ and\ \citenamefont
  {Raithel}}]{MacLennan2018}%
  \BibitemOpen
  \bibfield  {author} {\bibinfo {author} {\bibfnamefont {J.~L.}\ \bibnamefont
  {MacLennan}}, \bibinfo {author} {\bibfnamefont {Y.-J.}\ \bibnamefont {Chen}},
  \ and\ \bibinfo {author} {\bibfnamefont {G.}~\bibnamefont {Raithel}},\ }\href
  {https://arxiv.org/abs/1808.05136} {\bibfield  {journal} {\bibinfo  {journal}
  {arXiv:1808.05136}\ } (\bibinfo {year} {unpublished})}\BibitemShut {NoStop}%
\bibitem [{\citenamefont {Fabrikant}(1986)}]{Fabrikant1986}%
  \BibitemOpen
  \bibfield  {author} {\bibinfo {author} {\bibfnamefont {I.~I.}\ \bibnamefont
  {Fabrikant}},\ }\href
  {iopscience.iop.org/article/10.1088/0022-3700/19/10/021} {\bibfield
  {journal} {\bibinfo  {journal} {J. Phys. B}\ }\textbf {\bibinfo {volume}
  {19}},\ \bibinfo {pages} {1527} (\bibinfo {year} {1986})}\BibitemShut
  {NoStop}%
\bibitem [{\citenamefont {Bahrim}\ and\ \citenamefont
  {Thumm}(2000)}]{Bahrim2000}%
  \BibitemOpen
  \bibfield  {author} {\bibinfo {author} {\bibfnamefont {C.}~\bibnamefont
  {Bahrim}}\ and\ \bibinfo {author} {\bibfnamefont {U.}~\bibnamefont {Thumm}},\
  }\href {link.aps.org/pdf/10.1103/PhysRevA.61.022722} {\bibfield  {journal}
  {\bibinfo  {journal} {Phys. Rev. A}\ }\textbf {\bibinfo {volume} {61}},\
  \bibinfo {pages} {022722} (\bibinfo {year} {2000})}\BibitemShut {NoStop}%
\bibitem [{\citenamefont {Bahrim}\ \emph {et~al.}(2001)\citenamefont {Bahrim},
  \citenamefont {Thumm},\ and\ \citenamefont {Fabrikant}}]{Bahrim2001}%
  \BibitemOpen
  \bibfield  {author} {\bibinfo {author} {\bibfnamefont {C.}~\bibnamefont
  {Bahrim}}, \bibinfo {author} {\bibfnamefont {U.}~\bibnamefont {Thumm}}, \
  and\ \bibinfo {author} {\bibfnamefont {I.~I.}\ \bibnamefont {Fabrikant}},\
  }\href {https://link.aps.org/doi/10.1103/PhysRevA.63.042710} {\bibfield
  {journal} {\bibinfo  {journal} {Phys. Rev. A}\ }\textbf {\bibinfo {volume}
  {63}},\ \bibinfo {pages} {042710} (\bibinfo {year} {2001})}\BibitemShut
  {NoStop}%
\bibitem [{\citenamefont {Eiles}(2018)}]{Eiles2018}%
  \BibitemOpen
  \bibfield  {author} {\bibinfo {author} {\bibfnamefont {M.~T.}\ \bibnamefont
  {Eiles}},\ }\href {\doibase 10.1103/PhysRevA.98.042706} {\bibfield  {journal}
  {\bibinfo  {journal} {Phys. Rev. A}\ }\textbf {\bibinfo {volume} {98}},\
  \bibinfo {pages} {042706} (\bibinfo {year} {2018})}\BibitemShut {NoStop}%
\bibitem [{\citenamefont {Johnston}\ and\ \citenamefont
  {Burrow}(1982)}]{Johnston1982}%
  \BibitemOpen
  \bibfield  {author} {\bibinfo {author} {\bibfnamefont {A.~R.}\ \bibnamefont
  {Johnston}}\ and\ \bibinfo {author} {\bibfnamefont {P.~D.}\ \bibnamefont
  {Burrow}},\ }\href {iopscience.iop.org/article/10.1088/0022-3700/15/20/005/}
  {\bibfield  {journal} {\bibinfo  {journal} {J. Phys. B}\ }\textbf {\bibinfo
  {volume} {15}},\ \bibinfo {pages} {L745} (\bibinfo {year}
  {1982})}\BibitemShut {NoStop}%
\bibitem [{\citenamefont {Johnston}\ and\ \citenamefont
  {Burrow}(1995)}]{Johnston1995}%
  \BibitemOpen
  \bibfield  {author} {\bibinfo {author} {\bibfnamefont {A.~R.}\ \bibnamefont
  {Johnston}}\ and\ \bibinfo {author} {\bibfnamefont {P.~D.}\ \bibnamefont
  {Burrow}},\ }\href {https://link.aps.org/doi/10.1103/PhysRevA.51.406}
  {\bibfield  {journal} {\bibinfo  {journal} {Phys. Rev. A}\ }\textbf {\bibinfo
  {volume} {51}},\ \bibinfo {pages} {406} (\bibinfo {year} {1995})}\BibitemShut
  {NoStop}%
\bibitem [{\citenamefont {Scheer}\ \emph {et~al.}(1998)\citenamefont {Scheer},
  \citenamefont {Th{\o}gersen}, \citenamefont {Bilodeau}, \citenamefont
  {Brodie}, \citenamefont {Haugen}, \citenamefont {Andersen}, \citenamefont
  {Kristensen},\ and\ \citenamefont {Andersen}}]{Scheer1998}%
  \BibitemOpen
  \bibfield  {author} {\bibinfo {author} {\bibfnamefont {M.}~\bibnamefont
  {Scheer}}, \bibinfo {author} {\bibfnamefont {J.}~\bibnamefont
  {Th{\o}gersen}}, \bibinfo {author} {\bibfnamefont {R.~C.}\ \bibnamefont
  {Bilodeau}}, \bibinfo {author} {\bibfnamefont {C.~A.}\ \bibnamefont
  {Brodie}}, \bibinfo {author} {\bibfnamefont {H.~K.}\ \bibnamefont {Haugen}},
  \bibinfo {author} {\bibfnamefont {H.~H.}\ \bibnamefont {Andersen}}, \bibinfo
  {author} {\bibfnamefont {P.}~\bibnamefont {Kristensen}}, \ and\ \bibinfo
  {author} {\bibfnamefont {T.}~\bibnamefont {Andersen}},\ }\href
  {link.aps.org/pdf/10.1103/PhysRevLett.80.684} {\bibfield  {journal} {\bibinfo
   {journal} {Phys. Rev. Lett.}\ }\textbf {\bibinfo {volume} {80}},\ \bibinfo
  {pages} {684} (\bibinfo {year} {1998})}\BibitemShut {NoStop}%
\bibitem [{\citenamefont {Khuskivadze}\ \emph {et~al.}(2002)\citenamefont
  {Khuskivadze}, \citenamefont {Chibisov},\ and\ \citenamefont
  {Fabrikant}}]{Khuskivadze2002}%
  \BibitemOpen
  \bibfield  {author} {\bibinfo {author} {\bibfnamefont {A.~A.}\ \bibnamefont
  {Khuskivadze}}, \bibinfo {author} {\bibfnamefont {M.~I.}\ \bibnamefont
  {Chibisov}}, \ and\ \bibinfo {author} {\bibfnamefont {I.~I.}\ \bibnamefont
  {Fabrikant}},\ }\href {link.aps.org/pdf/10.1103/PhysRevA.66.042709}
  {\bibfield  {journal} {\bibinfo  {journal} {Phys. Rev. A}\ }\textbf {\bibinfo
  {volume} {66}},\ \bibinfo {pages} {042709} (\bibinfo {year}
  {2002})}\BibitemShut {NoStop}%
\bibitem [{\citenamefont {Markson}\ \emph {et~al.}(2016)\citenamefont
  {Markson}, \citenamefont {Rittenhouse}, \citenamefont {Schmidt},
  \citenamefont {Shaffer},\ and\ \citenamefont {Sadeghpour}}]{Markson2016}%
  \BibitemOpen
  \bibfield  {author} {\bibinfo {author} {\bibfnamefont {S.}~\bibnamefont
  {Markson}}, \bibinfo {author} {\bibfnamefont {S.~T.}\ \bibnamefont
  {Rittenhouse}}, \bibinfo {author} {\bibfnamefont {R.}~\bibnamefont
  {Schmidt}}, \bibinfo {author} {\bibfnamefont {J.~P.}\ \bibnamefont
  {Shaffer}}, \ and\ \bibinfo {author} {\bibfnamefont {H.~R.}\ \bibnamefont
  {Sadeghpour}},\ }\href@noop {} {\bibfield  {journal} {\bibinfo  {journal}
  {ChemPhysChem}\ }\textbf {\bibinfo {volume} {17}},\ \bibinfo {pages} {3683}
  (\bibinfo {year} {2016})}\BibitemShut {NoStop}%
\bibitem [{\citenamefont {Eiles}\ and\ \citenamefont
  {Greene}(2017)}]{Eiles2017}%
  \BibitemOpen
  \bibfield  {author} {\bibinfo {author} {\bibfnamefont {M.~T.}\ \bibnamefont
  {Eiles}}\ and\ \bibinfo {author} {\bibfnamefont {C.~H.}\ \bibnamefont
  {Greene}},\ }\href
  {https://journals.aps.org/pra/abstract/10.1103/PhysRevA.95.042515} {\bibfield
   {journal} {\bibinfo  {journal} {Phys. Rev. A}\ }\textbf {\bibinfo {volume}
  {95}},\ \bibinfo {pages} {042515} (\bibinfo {year} {2017})}\BibitemShut
  {NoStop}%
\bibitem [{\citenamefont {Hummel}\ \emph {et~al.}(2018)\citenamefont {Hummel},
  \citenamefont {Fey},\ and\ \citenamefont {Schmelcher}}]{Hummel2017}%
  \BibitemOpen
  \bibfield  {author} {\bibinfo {author} {\bibfnamefont {F.}~\bibnamefont
  {Hummel}}, \bibinfo {author} {\bibfnamefont {C.}~\bibnamefont {Fey}}, \ and\
  \bibinfo {author} {\bibfnamefont {P.}~\bibnamefont {Schmelcher}},\ }\href
  {https://link.aps.org/doi/10.1103/PhysRevA.97.043422} {\bibfield  {journal}
  {\bibinfo  {journal} {Phys. Rev. A}\ }\textbf {\bibinfo {volume} {97}},\
  \bibinfo {pages} {043422} (\bibinfo {year} {2018})}\BibitemShut {NoStop}%
\bibitem [{\citenamefont {Li}\ \emph {et~al.}(2003)\citenamefont {Li},
  \citenamefont {Mourachko}, \citenamefont {Noel},\ and\ \citenamefont
  {Gallagher}}]{Li2003}%
  \BibitemOpen
  \bibfield  {author} {\bibinfo {author} {\bibfnamefont {W.}~\bibnamefont
  {Li}}, \bibinfo {author} {\bibfnamefont {I.}~\bibnamefont {Mourachko}},
  \bibinfo {author} {\bibfnamefont {M.~W.}\ \bibnamefont {Noel}}, \ and\
  \bibinfo {author} {\bibfnamefont {T.~F.}\ \bibnamefont {Gallagher}},\ }\href
  {link.aps.org/pdf/10.1103/PhysRevA.67.052502} {\bibfield  {journal} {\bibinfo
   {journal} {Phys. Rev. A}\ }\textbf {\bibinfo {volume} {67}},\ \bibinfo
  {pages} {052502} (\bibinfo {year} {2003})}\BibitemShut {NoStop}%
\bibitem [{\citenamefont {Han}\ \emph {et~al.}(2006)\citenamefont {Han},
  \citenamefont {Jamil}, \citenamefont {Norum}, \citenamefont {Tanner},\ and\
  \citenamefont {Gallagher}}]{Han2006}%
  \BibitemOpen
  \bibfield  {author} {\bibinfo {author} {\bibfnamefont {J.}~\bibnamefont
  {Han}}, \bibinfo {author} {\bibfnamefont {Y.}~\bibnamefont {Jamil}}, \bibinfo
  {author} {\bibfnamefont {D.~V.~L.}\ \bibnamefont {Norum}}, \bibinfo {author}
  {\bibfnamefont {P.~J.}\ \bibnamefont {Tanner}}, \ and\ \bibinfo {author}
  {\bibfnamefont {T.~F.}\ \bibnamefont {Gallagher}},\ }\href
  {https://link.aps.org/doi/10.1103/PhysRevA.74.054502} {\bibfield  {journal}
  {\bibinfo  {journal} {Phys. Rev. A}\ }\textbf {\bibinfo {volume} {74}},\
  \bibinfo {pages} {054502} (\bibinfo {year} {2006})}\BibitemShut {NoStop}%
\bibitem [{\citenamefont {Arimondo}\ \emph {et~al.}(1977)\citenamefont
  {Arimondo}, \citenamefont {Inguscio},\ and\ \citenamefont
  {Violino}}]{Arimondo1977}%
  \BibitemOpen
  \bibfield  {author} {\bibinfo {author} {\bibfnamefont {E.}~\bibnamefont
  {Arimondo}}, \bibinfo {author} {\bibfnamefont {M.}~\bibnamefont {Inguscio}},
  \ and\ \bibinfo {author} {\bibfnamefont {P.}~\bibnamefont {Violino}},\ }\href
  {https://link.aps.org/doi/10.1103/RevModPhys.49.31} {\bibfield  {journal}
  {\bibinfo  {journal} {Rev. Mod. Phys.}\ }\textbf {\bibinfo {volume} {49}},\
  \bibinfo {pages} {31} (\bibinfo {year} {1977})}\BibitemShut {NoStop}%
\bibitem [{\citenamefont {Du}\ and\ \citenamefont {Greene}(1987)}]{Du1987}%
  \BibitemOpen
  \bibfield  {author} {\bibinfo {author} {\bibfnamefont {N.~Y.}\ \bibnamefont
  {Du}}\ and\ \bibinfo {author} {\bibfnamefont {C.~H.}\ \bibnamefont
  {Greene}},\ }\href {https://link.aps.org/doi/10.1103/PhysRevA.36.971}
  {\bibfield  {journal} {\bibinfo  {journal} {Phys. Rev. A}\ }\textbf {\bibinfo
  {volume} {36}},\ \bibinfo {pages} {971} (\bibinfo {year} {1987})}\BibitemShut
  {NoStop}%
\bibitem [{\citenamefont {Fey}\ \emph {et~al.}(2015)\citenamefont {Fey},
  \citenamefont {Kurz}, \citenamefont {Schmelcher}, \citenamefont
  {Rittenhouse},\ and\ \citenamefont {Sadeghpour}}]{Fey2015}%
  \BibitemOpen
  \bibfield  {author} {\bibinfo {author} {\bibfnamefont {C.}~\bibnamefont
  {Fey}}, \bibinfo {author} {\bibfnamefont {M.}~\bibnamefont {Kurz}}, \bibinfo
  {author} {\bibfnamefont {P.}~\bibnamefont {Schmelcher}}, \bibinfo {author}
  {\bibfnamefont {S.~T.}\ \bibnamefont {Rittenhouse}}, \ and\ \bibinfo {author}
  {\bibfnamefont {H.~R.}\ \bibnamefont {Sadeghpour}},\ }\href
  {http://iopscience.iop.org/article/10.1088/1367-2630/17/5/055010/meta}
  {\bibfield  {journal} {\bibinfo  {journal} {New J. Phys.}\ }\textbf {\bibinfo
  {volume} {17}},\ \bibinfo {pages} {055010} (\bibinfo {year}
  {2015})}\BibitemShut {NoStop}%
\bibitem [{\citenamefont {Tarana}\ and\ \citenamefont
  {Curik}(2016)}]{Tarana2016}%
  \BibitemOpen
  \bibfield  {author} {\bibinfo {author} {\bibfnamefont {M.}~\bibnamefont
  {Tarana}}\ and\ \bibinfo {author} {\bibfnamefont {R.}~\bibnamefont {Curik}},\
  }\href {https://link.aps.org/doi/10.1103/PhysRevA.93.012515} {\bibfield
  {journal} {\bibinfo  {journal} {Phys. Rev. A}\ }\textbf {\bibinfo {volume}
  {93}},\ \bibinfo {pages} {012515} (\bibinfo {year} {2016})}\BibitemShut
  {NoStop}%
\bibitem [{\citenamefont {Kleinbach}\ \emph {et~al.}(2017)\citenamefont
  {Kleinbach}, \citenamefont {Meinert}, \citenamefont {Engel}, \citenamefont
  {Kwon}, \citenamefont {L{\"{o}}w}, \citenamefont {Pfau},\ and\ \citenamefont
  {Raithel}}]{Kleinbach2017}%
  \BibitemOpen
  \bibfield  {author} {\bibinfo {author} {\bibfnamefont {K.~S.}\ \bibnamefont
  {Kleinbach}}, \bibinfo {author} {\bibfnamefont {F.}~\bibnamefont {Meinert}},
  \bibinfo {author} {\bibfnamefont {F.}~\bibnamefont {Engel}}, \bibinfo
  {author} {\bibfnamefont {W.~J.}\ \bibnamefont {Kwon}}, \bibinfo {author}
  {\bibfnamefont {R.}~\bibnamefont {L{\"{o}}w}}, \bibinfo {author}
  {\bibfnamefont {T.}~\bibnamefont {Pfau}}, \ and\ \bibinfo {author}
  {\bibfnamefont {G.}~\bibnamefont {Raithel}},\ }\href
  {https://link.aps.org/doi/10.1103/PhysRevLett.118.223001} {\bibfield
  {journal} {\bibinfo  {journal} {Phys. Rev. Lett.}\ }\textbf {\bibinfo
  {volume} {118}},\ \bibinfo {pages} {223001} (\bibinfo {year}
  {2017})}\BibitemShut {NoStop}%
\end{thebibliography}

\end{document}